\documentclass[longbibliography,10pt,aps,physrev,twocolumn,superscriptaddress,amssymb]{revtex4-2}
\usepackage[english]{babel}

\usepackage[colorinlistoftodos, color=green!40, prependcaption]{todonotes}

\usepackage[colorlinks=true,linkcolor=magenta,citecolor=magenta,urlcolor=magenta,linktocpage=True]{hyperref}
\usepackage{amsthm}
\usepackage{mathtools}
\usepackage{xcolor}
\usepackage{graphicx}
\usepackage{adjustbox}
\usepackage{placeins}
\usepackage[T1]{fontenc}
\usepackage{lipsum}
\usepackage{csquotes}
\usepackage{cleveref}

\definecolor{red}{rgb}{0,0,0}

\DeclarePairedDelimiter{\abs}{\lvert}{\rvert}
\DeclarePairedDelimiter{\bra}{\langle}{\rvert}
\DeclarePairedDelimiter{\ket}{\lvert}{\rangle}
\DeclarePairedDelimiter{\ex}{\langle}{\rangle}

\newcommand{\figref}[1]{\mbox{Fig.~\ref{#1}}}

\renewcommand{\eqref}[1]{\mbox{Eq.~(\ref{#1})}}

\newcommand{\be}{\begin{equation}}
\newcommand{\ee}{\end{equation}}
\newcommand{\bea}{\begin{eqnarray}}
\newcommand{\eea}{\end{eqnarray}}

\begin{document}
\title{Strongly-coupled non-Markovian waveguide QED with input-output HEOM}

\author{Neill Lambert}
\email{nwlambert@gmail.com}
\affiliation{RIKEN Center for Quantum Computing, RIKEN, Wakoshi, Saitama 351-0198, Japan}
\author{Yi-Te Huang}
\affiliation{Department of Physics, National Cheng Kung University, Tainan 701401, Taiwan}
\affiliation{Center for Quantum Frontiers of Research and Technology (QFort), Tainan 701401, Taiwan}
\author{Yueh-Nan Chen}
\affiliation{Department of Physics, National Cheng Kung University, Tainan 701401, Taiwan}
\affiliation{Center for Quantum Frontiers of Research and Technology (QFort), Tainan 701401, Taiwan}
\affiliation{Physics Division, National Center for Theoretical Sciences, Taipei 106319, Taiwan}
\author{Paul Menczel}
\email{paul@menczel.net}
\affiliation{RIKEN Center for Quantum Computing, RIKEN, Wakoshi, Saitama 351-0198, Japan}
\author{Franco Nori}
\affiliation{RIKEN Center for Quantum Computing, RIKEN, Wakoshi, Saitama 351-0198, Japan}
\affiliation{Physics Department, University of Michigan, Ann Arbor, MI 48109-1040, USA}

\date{\today} 

\begin{abstract}
We consider the problem of modeling a single qubit in contact with a one-dimensional waveguide beyond the standard perturbative and Markovian approximations. Using the recently developed input-output hierarchical equations of motion (io-HEOM), we investigate multiple examples of such waveguides, characterized by different spectral densities. Our examples highlight that the io-HEOM method can accurately capture non-Markovianity in waveguide QED from two distinct origins. The first source of non-Markovianity is  spatially non-local coupling between the qubit and the waveguide. By examining two examples with non-local coupling, we show how the coupling function affects the steady-state bound photons, and demonstrate the release of these photons when the qubit energy is quenched. The second source of non-Markovianity is non-linear dispersion. We illustrate this scenario using the example of a cavity array with point-like coupling, where the non-linear dispersion leads to persistent oscillations due to Van Hove singularities in the spectral density. 
\end{abstract}


\maketitle

\section{Introduction}

Ultra-strong coupling in cavity QED~\cite{Kockum18b,forn2018ultrastrong} has led to a range of exciting physical predictions: the potential to observe virtual excitations in a laboratory setting~\cite{de2017virtual, Stassi13}, the emergence of ground-state electroluminescence from these excitations~\cite{Cirio16,PhysRevLett.122.190403}, and complex processes that are forbidden or impossible in standard quantum optics \cite{Kuzmin2019}. With the use of superconducting circuits \cite{Yoshihara2017} to realize these effects, it is also clear that one is not confined to zero-dimensional cavity limits, but can also reach ultra-strong coupling with extended systems such as one-dimensional waveguides and cavity arrays \cite{PhysRevLett.120.140404, Kannan2020, PhysRevA.99.032325, goran, niemczyk2010circuit, phonons, magazzu2018probing, Puertas, entharve, gong, zhou1, helmut1, helmut2}.

\begin{figure}[!t]
\includegraphics[width=\columnwidth]{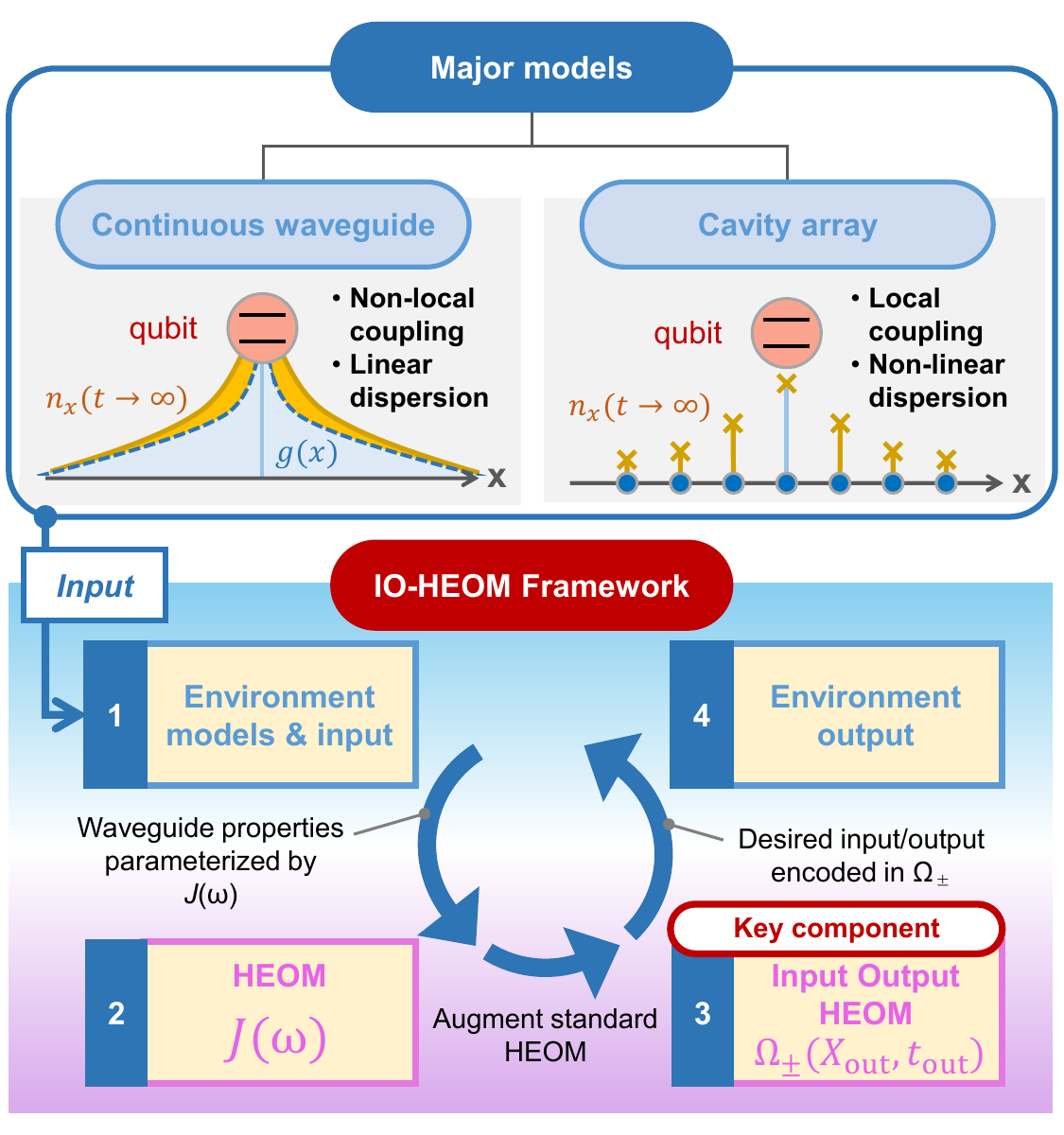}
\caption{
In this work we study two examples of non-Markovian waveguide QED: a qubit coupled either to a continous waveguide with non-local coupling (the blue shaded region) and linear dispersion (top left), or to a cavity array with local coupling (blue line) and non-linear dispersion (top right).  By casting both cases into a standard open quantum systems framework, via the spectral density $J(\omega)$, we can treat these cases with the hierarchical equations of motion. To recover how the waveguide photon occupation $n_x(t)$ (yellow shaded region in the upper left figure, yellow crosses in the upper right figure) responds due to the interaction with the qubit we extend the recently developed input-output HEOM (io-HEOM) framework, which correctly predicts the non-Markovian dynamics and bound-photon steady states (depicted schematically in the figure) of both scenarios with one unified method.
}\label{schematic}    
\end{figure}

Many aspects of the physics of the ultra-strong coupling regime in waveguide QED have been explored theoretically, including bound-state photons \cite{zuecoold, zuecodynamicbound}, their emission from quenches \cite{zueco_quench}, and the role of gauge invariance \cite{zuecodipolegauge, PhysRevA.98.053819}. These studies employ several powerful numerical and analytical techniques, including numerical simulations using matrix product states \cite{zuecoold,stephenhughes1, SanchezMunoz2018} (particularly amenable to cavity array models, see \figref{schematic}),  and variational polaron methods.  With these tools, it has been shown that bound-state photons can appear in the waveguide around the qubit, both because of ultra-strong coupling and because of band gaps \cite{zuecodynamicbound}. They have also been used to demonstrate the transition of bound-state photons from virtual to real with non-adiabatic quenches of qubit properties \cite{zueco_quench}.

However, a full exact numerical method valid for all coupling regimes, time scales, and dimensionalities is, so far, missing.  Techniques combining discrete waveguide models with matrix product state techniques are incredibly powerful, but often limited to finite times and one spatial dimension. Reaching the spatial continuum and the temporal steady state limit is challenging, and matrix product state methods can become inefficient as dimensionality is increased~\cite{Tay2025}. Here, we demonstrate that the powerful hierarchical equations of motion (HEOM) technique \cite{Tanimura_1,Tanimura_2,Tanimura_3,Tanimura_2020, Lambert_Bofin} can be used for these challenging regimes by harnessing an extended version of the hierarchical equations that incorporate input and output fields \cite{ioheom}. This allows for the efficient reconstruction of waveguide properties and for the simulation of scattering experiments. 

To demonstrate the effectivity of the HEOM technique, we apply it to waveguide QED problems featuring non-Markovian \cite{zhangprl, RevModPhys.89.015001, Lidar, Petruccione} and non-perturbative \cite{Lambert} effects, which arise from two distinct origins (see \figref{schematic}).

First, in the limit of a continuous waveguide with linear dispersion, non-Markovianity arises from non-local coupling functions \cite{zuecogiant, goran, continouscoupl}. We focus on two examples for this case, one where the coupling function gives rise to an effective Ohmic bath with an exponential cutoff, and one where the effective bath has a Lorentzian spectral density. We study the dependence of the bound-state photon density in the long-time limit on said non-local coupling function, and verify our results with a variational polaron calculation.  For the Ohmic case, we also demonstrate the emission pulse that occurs when a qubit is non-adiabatically quenched above the high-frequency cut-off, easily simulatable with the HEOM. 

Second, we show that non-Markovian and non-perturbative effects can also arise when the qubit-waveguide coupling is point-like, but the waveguide itself has non-linear dispersion.  To demonstrate this effect, we focus on the example of a cavity array, relevant for coupled-resonator waveguides, microwave photonic crystals, and Josephson metamaterial waveguides \cite{painter,gasparinetti}. The cavity array model features slowly decaying free correlation functions that are challenging to capture with the HEOM, and lead to more dynamic non-Markovian effects, in addition to bound-state photons.  We show that modern correlation function fitting techniques, such as the Prony method \cite{prony2,prony1,prony3}, alleviate these challenges to some degree. The input-output HEOM approach therefore allows us to study both scenarios, non-local coupling and non-linear dispersion, on an equal footing with one unified method. A Github repository is provided as supplementary information for reproducing all examples in this paper, as well as animated versions of the space-time plots \cite{code}.

\section{Waveguide QED}

We model a one-dimensional waveguide, interacting with a qubit, with the continuum Hamiltonian,
\begin{align}
    H_{\mathrm{tot}} &= H_S +  \int_{-\infty}^\infty dp\; \bar\omega(p) b^\dagger(p) b(p) \nonumber\\
    &\quad + \sigma_x\int_{-\infty}^\infty dp \; g(p) \bigl[b(p)+b^\dagger(p)\bigr]\;,
\label{eq:1D}
\end{align}
where $H_S = (\omega_s /2)\sigma_z$,  
$\bar \omega(p)$ and $g(p)$ are the dispersion relation and the coupling function, and $b(p)$ are annihilation operators for waveguide modes with $[b(p),b(p')]=\delta(p-p')$. The momentum basis can be transformed to the position basis with 
\begin{equation}
b(x)=\frac{1}{\sqrt{2\pi}}\int_{-\infty}^\infty dp\; e^{i p x} b(p)\;.\label{ft}
\end{equation}

In general, the dispersion and coupling in momentum space can be related to the traditional open-system concept of spectral density by a change of variables, 
\begin{equation}
J_{i}(\omega) = \pi\, \abs[\big]{g(p_i^\omega)}^2\,\abs[\bigg]{\frac{d\bar\omega}{dp}(p_i^\omega)}^{-1}\;, \label{jfund}
\end{equation}
where $p_i^\omega$ are the (potentially degenerate) solutions of $\bar\omega(p)=\omega$, and the notation in the denominator stands for the derivative of $\bar\omega(p)$ evaluated at $p_i^\omega$.

If we assume linear dispersion [$\bar\omega(p) = c |p|$] and point-like coupling in space [$g(x) \propto \delta(x-x_0)$], \eqref{jfund} implies that the waveguide is Markovian in nature since the spectral density becomes constant.
To obtain a structured spectral density and non-Markovian behavior, we thus need non-linear dispersion, non-local coupling, or both (note that for our purposes other sources of high frequency cut-off can be absorbed into an effective non-local coupling function). The input-output HEOM method \cite{ioheom}, which we introduce below, has already been successfully applied to the Markovian case in \cite{ioheom}. In this paper, we demonstrate that the io-HEOM method can also accurately non-Markovian effects arising from non-local coupling or non-linear dispersion, and give insight into dynamical behavior of the waveguide itself (which is normally unobtainable with standard HEOM).

Note, for completeness, that another type of non-Markovianity can arise, even in the linear dispersion and local coupling limits, when time-delay \cite{liang2024purifiedinputoutputpseudomodemodel} is involved (e.g., via ``mirrors'') \cite{zhou1, zhou2, stephenhughes1, stephenhughes2, Hoi2015}.  However we don't consider this case in this work.

\subsection{Brief Summary of the io-HEOM Framework}

Before moving on to the explicit examples, we briefly summarize the io-HEOM formalism. The HEOM method \cite{Tanimura_1,Tanimura_2,Tanimura_3,Tanimura_2020, Lambert_Bofin} is traditionally used to simulate non-Markovian open quantum systems exactly. Given a system Hamiltonian, a bath spectral density and temperature, and a system operator that couples to the bath, the HEOM is a set of coupled equations of motion for the reduced system operator $\rho^{(0)}(t)$ and a set of auxiliary density operators (ADOs) $\rho^{(\bar n)}(t)$, where $\bar n$ is a multi-index. As a shorthand for these equations, we write
\begin{equation} \label{eq:heom_short}
    \dot\rho^{(\bar{n})}(t) = \mathcal{M}_{\mathrm{HEOM}}\Bigl[ \rho^{(\bar{n})}(t) \Bigr] \;,
\end{equation}
the full expression can be found in \eqref{heomeom} in Appendix~\ref{app:heom}.  The HEOM method relies on a multi-exponential ansatz [see \eqref{ansatz}] for the free bath correlation function [defined in \eqref{ct}], and its accuracy depends primarily on how well this ansatz fits the true bath correlation function.

While it has been shown that certain bath information, such as the particle \cite{Yan_5} or heat currents \cite{10.1063/1.4971370} exchanged with the baths, can be extracted from the ADOs, computing arbitrary bath observables is challenging with the standard HEOM method. Similarly, using non-thermal and in particular non-Gaussian bath initial states is also difficult.

The recently developed input-output HEOM (io-HEOM) theory \cite{ioheom} lifts both these restrictions by augmenting the standard HEOM with additional ADOs that are associated with the required input or output fields:
\begin{align}
\dot{\rho}^{(\bar{n}^{\phi},\bar{n})}(t) &= \mathcal{M}_{\mathrm{HEOM}}\Bigl[ \rho^{(\bar{n}^{\phi},\bar{n})}(t) \Bigr] \nonumber\\
&\quad + \sum_{j} n_j^{\phi}\, \mathcal{Y}_j(\tau_j,t)\, \rho^{(\bar{n}_j^{\phi}-1,\bar{n})}(t).\label{ioheom}
\end{align}
As with the standard HEOM, $\rho^{(0,0)}(t)$ is the reduced system state, and the multi-index $\bar{n}$ is related to the terms in the multi-exponential ansatz for the bath correlation function. The first term on the right hand side is identical to the standard HEOM, \eqref{eq:heom_short}.
However, in \eqref{ioheom},  the multi-index has been extended to the tuple $(\bar{n}^{\phi},\bar{n})$, where $\bar{n}^{\phi}=(n_0^{\phi},\cdots,n_{N_j}^{\phi})$ with $n_j^{\phi} \in (0,1)$ is an additional multi-index. Further, $N_j$ is the number of operators applied to the bath, $\tau_j$ the times where they are applied, and $\rho^{(\bar{n}_j^{\phi}-1,\bar{n})}$ refers to the ADO with the $j$-th entry of the multi-index $\bar{n}^{\phi}$ reduced by one. 
 
The action of the input-output fields acting on the environment is encoded in the time-dependent super-operators
\begin{equation} \label{eq:Yj_taut}
\mathcal{Y}_j(\tau_j, t) = \ex[\big]{\phi_j(\tau_j)\chi^l(t)}\, \mathcal{S}^l(t) - \ex[\big]{\phi_j(\tau_j)\chi^r(t)}\, \mathcal{S}^r(t) \;.
\end{equation}
They are composed of the system superoperators $\mathcal{S}^{l}(t)\rho = S(t)\rho$ and $\mathcal{S}^{r}(t)\rho = \rho S(t)$, where $S(t)$ is the system coupling operator in the interaction picture, and of the time-ordered free bath correlation functions $\ex{\phi_j(\tau_j)\chi^{l,r}(t)}$ between the applied fields $\phi_j(\tau_j)$ and the bath coupling operator. The notation $\chi^{l,r}(t)$ again denotes superoperators following the rule $\chi^{l}(t)\rho= X(t)\rho$ and $\chi^{r}(t)\rho= \rho X(t)$, where $X(t)$ is the bath coupling operator defined in Appendix~\ref{app:heom}.

We still have to relate the extended set of ADOs in the io-HEOM framework to the concrete quantities of interest in an input-output problem. In general, these quantities have the general form of conditioned bath correlation in the interaction picture,
\begin{equation}
\Phi_{N_j}(\tau_j, t) = \mathrm{Tr}\Biggl[ \mathcal{T} \prod_{j=1}^{N_j} \phi_j(\tau_j)\rho(t) \Biggr] \;, \label{phi}
\end{equation}
where $\rho(t)$  is the state of the full system and environment, and $\phi_j(\tau_j)$ are superoperators acting on the environment Hilbert space.  

In the general theory, the fields can either be ``static'' or ``dynamical'', meaning that $\tau_j$ is either fixed or equal to the dynamical time $t$.
In input theory, we consider static fields with $\tau_j = 0$ modifying the initial state.
Output theory can be realized in two equivalent ways: using dynamical fields, or static fields where $\tau_j = t_\text{out}$ is the output time, which is varied in a separate step, requiring the full simulation to be repeated for each considered value of $t_\text{out}$.

In the case of static fields, the conditioned correlations \eqref{phi} can be written in terms of the additional ADOs in the extended HEOM \eqref{ioheom}.
For the full general relation between the conditioned correlations and the additional ADOs, as well as for the more complicated extended HEOM in the case of dynamical fields, we refer to \cite{ioheom}.
In this work, we focus on the output theory of quadratic bath observables; that is, on computing waveguide properties such as the photon density at the final observation time $t_\text{out}$.
In this case, the relation becomes
\begin{align}
\Phi_2(t_\text{out}) &= \mathrm{Tr}\bigl[ \phi_1(t_\text{out})\phi_2(t_\text{out})\rho(t_\text{out}) \bigr] \nonumber \\
&= \ex[\big]{ \phi_1(t_\text{out})\phi_2(t_\text{out}) } - \mathrm{Tr}_S\bigl[ \rho^{((1,1),0)}(t_\text{out}) \bigr] \;,
\end{align}
where $\rho^{((1,1),0)}(t)$ denotes the ADO with $n_{1}^{\phi}=1$, $n_2^{\phi}=1$ and $\bar n = 0$.
Studying the role of input fields, for example via pulse scattering \cite{PhysRevLett.123.123604, theo} or continuous driving, will be reserved for future work.

A note should be made that these types of correlation functions can also be obtained from an alternative approach outlined in detail in \cite{Gribben2022usingenvironmentto}, which ultimately results in expressions of bath properties in terms of integrals over system two-time correlation functions.

\subsection{Waveguide Output Theory with io-HEOM}
Following \eqref{eq:1D}, the system and bath coupling operators for the waveguide, in the interaction picture, are 
\begin{equation}
S(t) = \sigma_- e^{-i\omega_s t}+\sigma_+ e^{i\omega_s t}
\end{equation}
and
\begin{equation}
X(t) = \int_{-\infty}^{\infty} dp\; g(p) \bigl[ b(p)e^{-i \bar\omega(p)t} + b(p)^{\dagger}e^{i \bar\omega(p)t} \bigr]\;, \label{eq:Xt}
\end{equation}
respectively.
For the HEOM description of the system-waveguide interaction, we follow the standard approach of defining the spectral density \eqref{jfund} and decomposing the resulting free bath correlation functions into exponential terms, see \eqref{ct}.

To compute the output of the waveguide using the input-output HEOM theory, we need to pre-compute the additional correlation functions appearing in \eqref{eq:Yj_taut}.
To proceed, we specify that the superoperators $\phi_j(t_\text{out})$ have the form $\phi_j(t_\text{out}) \rho = \varphi_j(t_\text{out}) \rho$, where $\varphi_j(t_\text{out})$ are the regular operators
\begin{align}
    \varphi_{1}(t_\text{out})
    &= \int_{-\infty}^\infty dp \;g_\text{out}(p) b^\dagger(p)e^{i\bar\omega(p) t_\text{out}} \quad \text{and} \nonumber\\
    \varphi_{2}(t_\text{out})
    &= \int_{-\infty}^\infty dp \;\bar{g}_\text{out}(p) b(p)e^{-i\bar\omega(p) t_\text{out}}\;. \label{eq:varphi-fields}
\end{align}
Note that more general constructions are possible, but this choice is convenient for our purposes here. Assuming that the waveguide is at zero temperature and initially in a vacuum state,
\eqref{eq:Yj_taut} then simplifies to
\begin{align}
    \mathcal{Y}_1(t_\text{out}, t) &= \ex[\big]{ X^-_t\varphi_1(t_\text{out}) }_\infty \bigl[ \theta(t - t_\text{out}) S(t) [\cdot] - [\cdot] S(t) \bigr] \\
\shortintertext{and}
    \mathcal{Y}_2(t_\text{out}, t) &= \ex[\big]{ \varphi_2(t_\text{out}) X^+_t }_\infty \theta(t_\text{out} - t) S(t)[\cdot] \;,
\end{align}
where $X_t^-$ and $X_t^+$ denotes the annihilator-part and the creator-part of $X(t)$ with $X(t) = X_t^- + X_t^+$. Further, $\ex{\cdot}_\infty$ is the zero-temperature ($\beta = \infty$) expectation value, $\theta$ the Heaviside step function, and $[\cdot]$ stands for the argument of the superoperators.
Note that in our examples, we will only consider times $t < t_\text{out}$.
   
In order to see the impact of the non-perturbative coupling with the qubit on the waveguide, we focus on looking at output that corresponds to the photon occupation in a region $\Delta x$ around the point $x_\text{out}$,
\begin{equation}
n(x_\text{out})=\Delta x\, b^\dagger(x_\text{out}) b(x_\text{out})\;.\label{output}
\end{equation}
We therefore set
\begin{align}
    \varphi_{1}(t_\text{out}) &= \sqrt{\Delta x}\, b^{\dagger}(x_{\text{out}},t_{\text{out}}) \quad \text{and} \nonumber\\
    \varphi_{2}(t_\text{out}) &= \sqrt{\Delta x}\, b(x_{\text{out}},t_{\text{out}}) \;.
\end{align}
Comparing with \eqref{eq:varphi-fields}, and using the Fourier relationship between $b(x)$ and $b(p)$ defined earlier in \eqref{ft},
we read off $g_{\text{out}}(p) = e^{-i p x_\text{out}}\sqrt{\Delta x/2 \pi}$.
With this choice, the additional correlation functions we need to construct $\mathcal{Y}_j(t_\text{out}, t)$ are given by
\begin{align}
   \Omega_+ &= \ex[\big]{ X^-_t\varphi_1(t_\text{out}) }_\infty \nonumber\\
   &= \sqrt{\frac{\Delta x}{2\pi}} \int_{-\infty}^{\infty}dp\; g(p)\;e^{-ipx_\text{out}}e^{i\bar\omega(p)(t_\text{out} - t)}
   \label{om_p} \\
\shortintertext{and}
    \Omega_- &= \ex[\big]{ \varphi_2(t_\text{out}) X^+_t }_\infty \nonumber\\ 
    &= \sqrt{\frac{\Delta x}{2\pi}} \int_{-\infty}^{\infty}dp\; g(p)e^{-ipx_\text{out}}e^{-i\bar\omega(p)(t_\text{out} - t)} \;.\label{om_m}
\end{align}
These correlation functions $\Omega_\pm$ essentially correspond to temporal-spatial correlation functions of the free waveguide filtered by the system-environment coupling function $g(p)$.

To summarize, in order to construct the input-output HEOM in \eqref{ioheom}, we first decompose or fit the free bath correlation function \eqref{ct} with exponential functions. We then compute the time-dependent correlation functions between the applied input or output fields and the bath coupling operator \eqref{eq:Xt}. This can often be done analytically, or numerically if need be. The results are inserted into the coupled time-dependent ODE \eqref{ioheom}, which is numerically integrated up to the output time $t_\text{out}$. In our case, the output time is the time when the output observable, \eqref{output}, is ``measured''.

\section{Example 1: Ohmic spectral density and linear dispersion}

Recall from \eqref{jfund} that the dispersion sets the relationship between the coupling function $g(p)$ and the spectral density $J(\omega)$. For our first example, we choose linear dispersion, $\bar\omega(p) = c |p|$, where the relationship is particularly simple.
Because each energy is associated with two degenerate momentum values, the system effectively sees two independent environments associated with left and right moving modes.
Since $g(p)$ is an even function, the two corresponding spectral densities $J_L(\omega) = J_R(\omega)$ are equal. The total spectral density felt by the system is
\begin{equation} \label{eq:linear_disp}
J(\omega) = J_L(\omega) + J_R(\omega) = \frac{2\pi}{c} \abs[\big]{ g\left(\pm \frac{\omega}{c}\right) }^2 \;.
\end{equation}

In this example, we choose to realize an Ohmic spectral density with exponential cut-off,
\begin{equation} \label{eq:ohmic}
J(\omega) = \lambda \omega \exp(-\omega/\omega_c) \;,
\end{equation}
where $\lambda$ is the coupling strength and $\omega_c$ the cutoff frequency, and we ask what continuous coupling function leads to this spectral density.
Solving \eqref{eq:linear_disp} gives $g(p) = \sqrt{c\, J(c|p|)/(2\pi)}$, and in position space
\begin{align}
    g(x) &= \frac{1}{\sqrt{2\pi}}\int_{-\infty}^\infty dp\; e^{i p x} \sqrt{c\, J(c|p|)/(2\pi)} \nonumber\\
        &=   \frac{c}{2}\sqrt{\frac{\lambda}{\pi}}\mathcal{R}\Bigl(\frac{c}{2\omega_c}-i x\Bigr)^{-{3/2}} \;,
\end{align}
where $\mathcal{R}$ means taking the real part. We see that the spatial coupling function in this case corresponds to a power law decay $|x|^{-3/2}$.  Typically, this coupling function should be seen as the microscopic entity which determines an effective spectral density (in this case Ohmic with exponential cut-off), and we can think of designing a continous coupling function as a type of environment engineering. Note that this continuous coupling function scenario \cite{continouscoupl} is akin, but slightly different to, standard giant atoms \cite{giantatoms, Kannan2020}, wherein a few well-separated coupling points lead to oscillatory spectral densities. 

To model this situation with the HEOM, we need the auto-correlation function of the waveguide. For this example, at zero temperature, it is
\begin{equation}
C(t) = \ex{ X(t) X(0) } = \frac{\lambda}{\pi} \omega_c^2\, (1+ i \omega_{c} t)^{-2} \;.
\end{equation}
The HEOM requires a multi-exponential decomposition of this function, which we obtain using standard fitting routines. 

To apply the extended io-HEOM and obtain output observables for the waveguide, we further need to evaluate the correlation functions between output and coupling terms, \eqref{om_p} and \eqref{om_m}.
We obtain
\begin{align}
\Omega_+  = \frac{\sqrt{\Delta x}}{2\pi} \int_{-\infty}^{\infty}dp\; c\,\sqrt{\lambda |p|}\, \exp \bigg\{ & -\frac{c|p|}{2\omega_c} -ipx_\text{out} \nonumber\\ &+ ic|p|(t_\text{out} - t)\bigg\}
\end{align}
and $\Omega_- = \Omega_+^\ast$.
This expression has the shape of a dressed Gamma function, which we can evaluate analytically to give
\begin{align}
\Omega_+ = \sqrt{\frac{\lambda\omega_c^{3}\, \Delta x}{2\pi c}} \biggl\{&\Bigl[1-2i\omega_c \Bigl(t_\text{out} - t -\frac{x_\text{out}}{c} \Bigr)\Bigr]^{-3/2} \nonumber\\
+ &\Bigl[1-2i\omega_c\Bigl(t_\text{out} - t + \frac{x_\text{out}}{c} \Bigr)\Bigr]^{-3/2} \biggr\} \; . 
\end{align}
No fitting is required for this function; it is used directly in the io-HEOM formalism.

\begin{figure}[!t]
\includegraphics[width=\columnwidth]{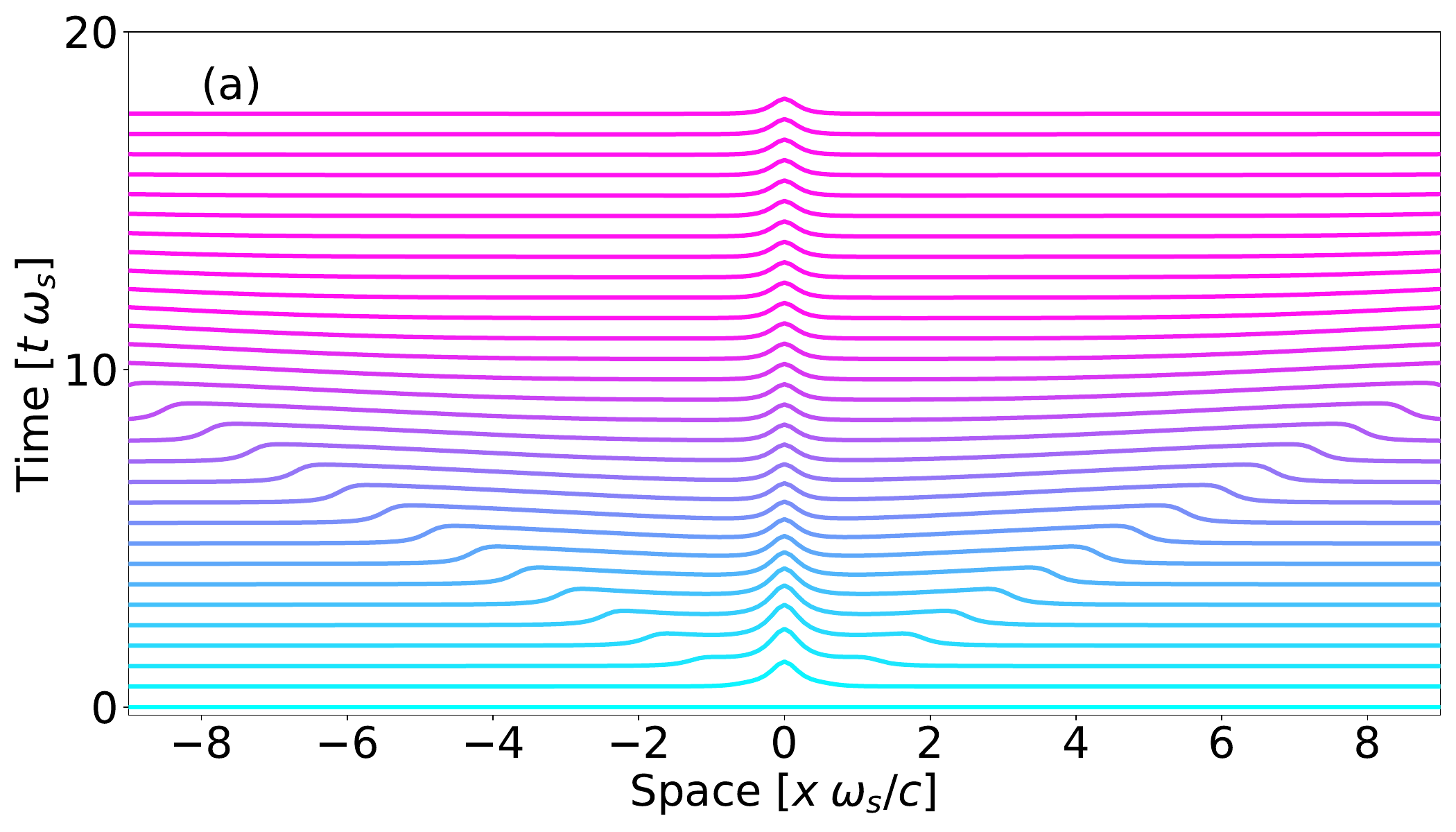}
\includegraphics[width=0.49\columnwidth]{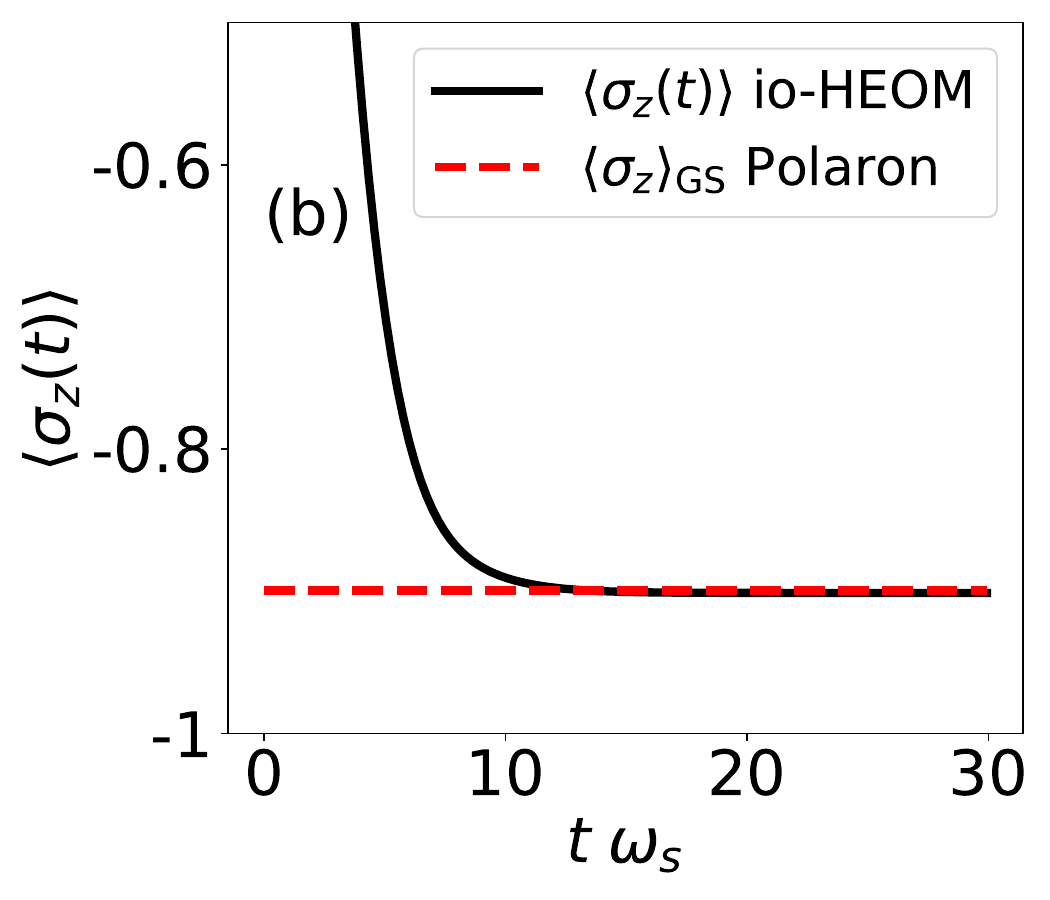}
\includegraphics[width=0.49\columnwidth]{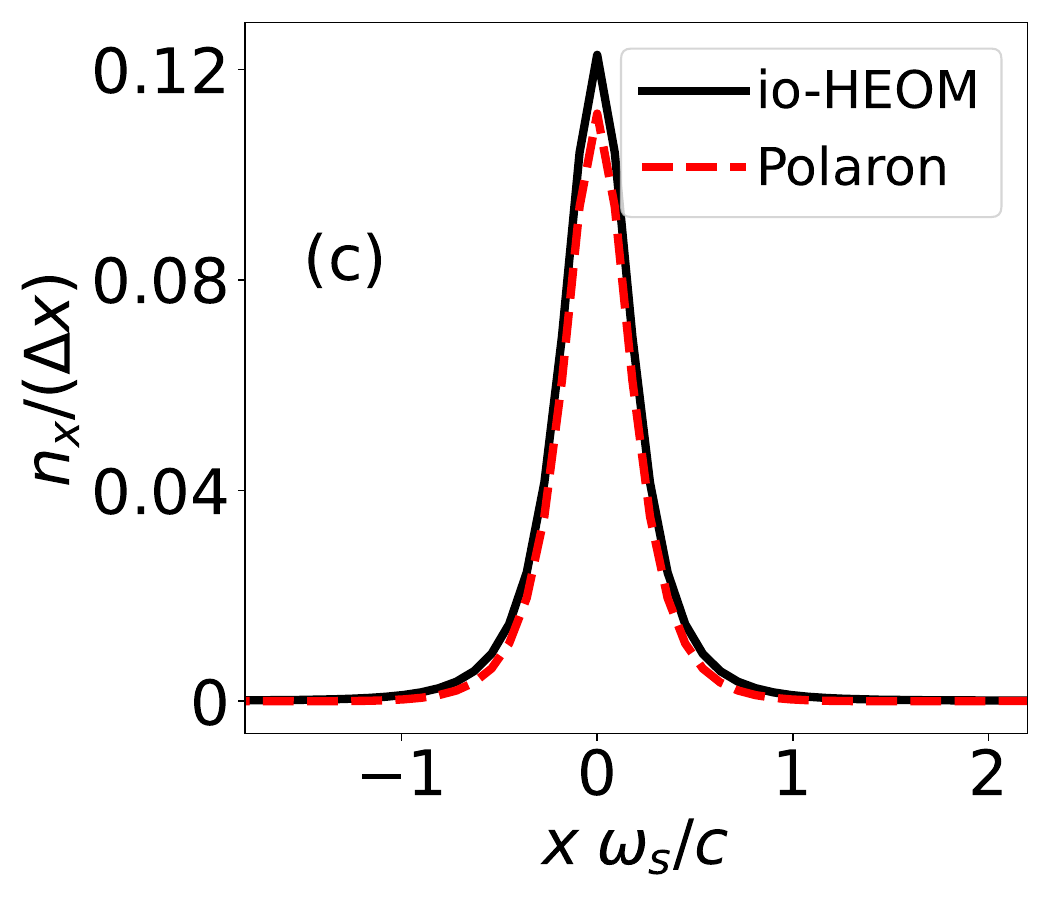}
\caption{
Dynamics and steady state in Example 1, an Ohmic waveguide with linear dispersion and exponential cut-off.
(a) We show the photon density $n_x(t)/\Delta x$ as a function of position $x \omega_s /c$, offset on the y-axis to represent different times $t \omega_s$. The lines are colored to make them easier to distinguish visually.
The spatial extent of the waveguide is in principle infinite; we only show a finite region here.
We use the Ohmic spectral density \eqref{eq:ohmic} with coupling strength $\lambda = 0.39$, exponential cutoff $\omega_c = 2\omega_s$, and zero temperature ($T=0$).  The initial condition is the waveguide in the vacuum and the qubit, at position $x=0$, excited.  We see two main features: a propagating wave front, moving left and right, and a residual bound photon occupation around the qubit that persists into the steady state.
We can, approximately, reproduce both (b) the residual qubit population and (c) the waveguide spatial profile in the steady state with Silbey-Harris polaron theory.
}\label{fig:ohmic}    
\end{figure}

\subsection{Emission dynamics and steady-state virtual photons}
\begin{figure*}[!t]
\includegraphics[width=1.33\columnwidth]{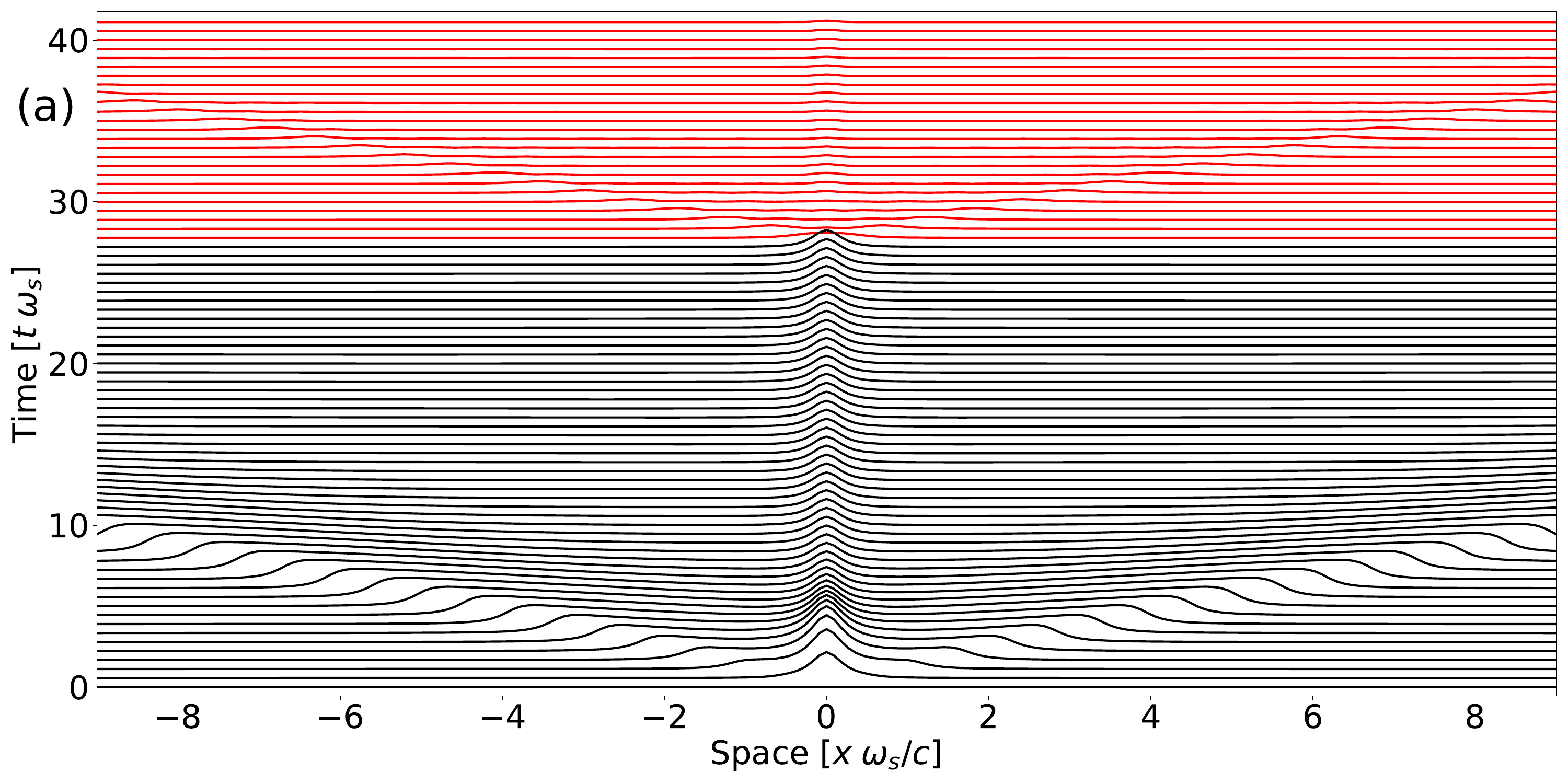}
\includegraphics[width=0.66\columnwidth]{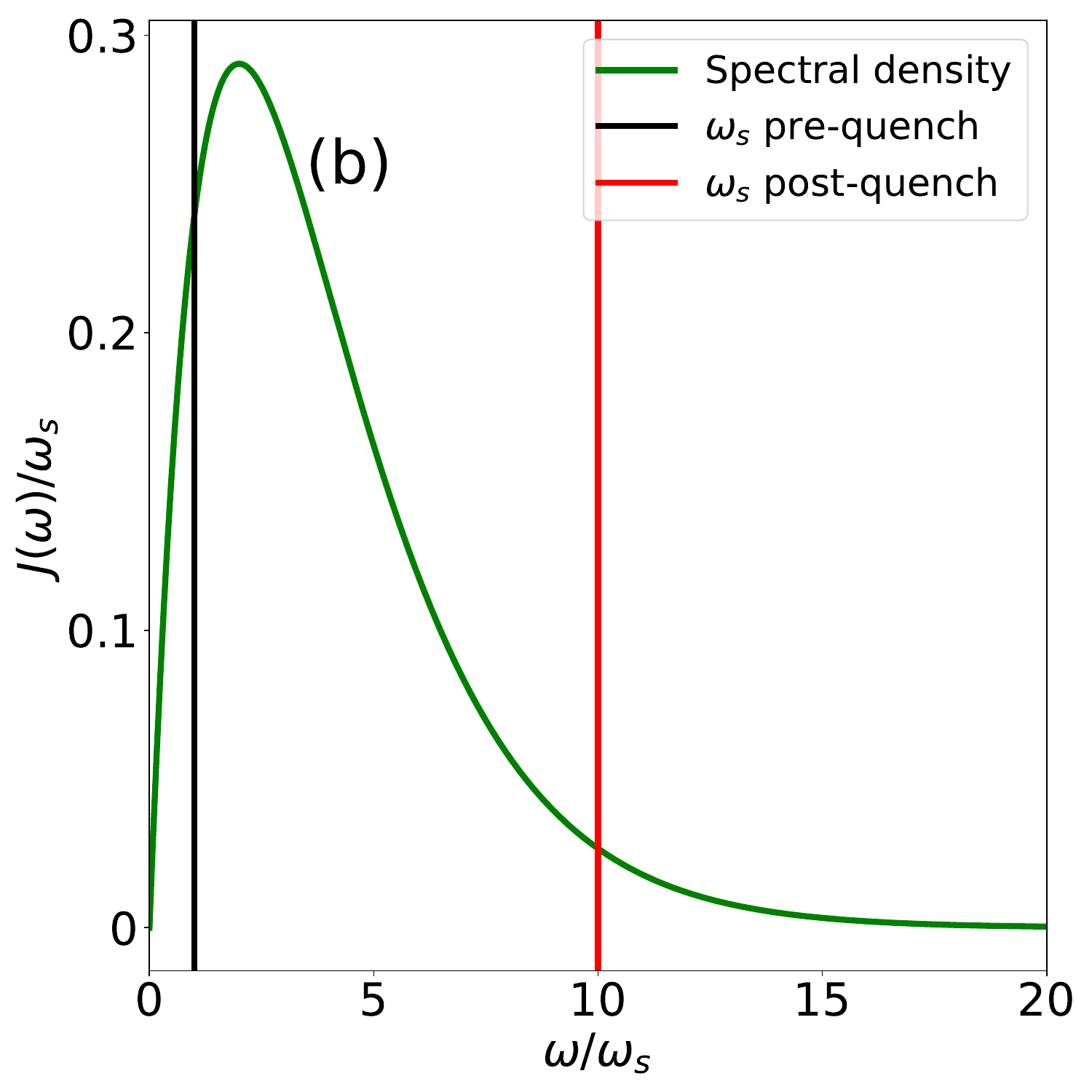}
\caption{
We demonstrate how a quench of the qubit frequency can lead to emission of photons out of the (virtual) bound state, for the case when a qubit is coupled to an Ohmic waveguide with exponential cut-off and linear dispersion.
(a) Photon density $n_x(t)/\Delta x$  as a function of position $x \omega_s /c$, offset on the y-axis to represent different times $t \omega_s$.
The color of the curves distinguishes pre-quench (black) and post-quench (red) times.
(b) The green curve shows the spectral density used in this example, an Ohmic spectral density with the same parameters as in \figref{fig:ohmic}. The vertical black line shows the qubit frequency pre-quench, and the vertical red line shows the frequency post-quench. As before, the initial condition is the waveguide in the vacuum and the qubit, at position $x=0$, excited.  After reaching the approximate steady-state, we non-adiabatically quench the qubit frequency so that the coupling to the waveguide is effectively weaker. This leads to an emission of photons towards the left and right, out of the bound state.
}\label{fig:ohmic_quench}    
\end{figure*}

When the interaction between the qubit and the waveguide is strong, one expects to observe a steady state where the system and the waveguide modes near it are hybridized, and the waveguide modes contain ``virtual'' excitations that are bound and cannot propagate. 
To demonstrate that the io-HEOM method can capture this effect, we prepare the qubit in an excited state and the waveguide in the vacuum, and observe the dynamics until a steady state has been reached (dynamics subsist).

Our simulation results in \figref{fig:ohmic} show the expected emission of an initial pulse, followed by the persistence of a small number of bound photons around the qubit position $x=0$. The system dynamics themselves do not exhibit clear non-Markovian effects for this choice of spectral density, but do show a subtly modified steady-state population associated with hybridization with the bound-state photons. 

As discussed earlier, the choice of spectral density determines the non-local spatial coupling between the qubit and the waveguide, described by $g(x)$.  This spatial profile can be thought of as a zero-order prediction for the bound photon density that we observe numerically. To get a more accurate analytical prediction, we employ a standard variational polaron method to recover the approximate ground state of the qubit-waveguide system. 
We change to a basis of coherent states in the waveguide, where each mode $b(p)$ is displaced by the qubit by parameters $f(p)$. The $f(p)$ displacements are found by variationally minimizing the energy.  We largely follow the steps in \cite{zueco_quench, Silbey}, which lead to a  renormalized qubit frequency
\begin{equation}
\omega_r = \omega_s \exp\biggl[-\frac{2}{\pi}\int_0^{\infty} d\omega\; \frac{J(\omega)}{(\omega+\omega_r)^2}\biggr] \label{omega_r}
\end{equation}
For the Ohmic bath with exponential cut-off, the integral evaluates to
\begin{equation}
\int_0^{\infty} d\omega\;  \frac{J(\omega)}{(\omega+\omega_r)^2} = \lambda \bigl[ (1+z)e^{z}E_1(z) - 1 \bigr] \;,
\end{equation}
where $z = \omega_r / \omega_c$ and $E_1(z) = \int_{z}^\infty dx\, \exp (-x)/x$.
We can then solve for $\omega_r$ self-consistently. As in the example in \cite{zueco_quench}, the expected qubit population in the ground-state polaron approximation is simply $\langle \sigma_z \rangle_{\mathrm{GS}} = -\omega_r/\omega_s$.  

Also the photon density at position $x$ in the polaron ground state, $n_x^{\mathrm{GS}} = \ex{b^{\dagger}(x)b(x)}_{\mathrm{GS}}$, can be found from the variational parameters. To this end, we continue following the steps in \cite{zueco_quench, Silbey} and express $b(x)$ in terms of $b(p)$, and then insert the unitary transformation that diagonalizes the original problem into the polaron basis.  The resulting expression simplifies drastically, and one finds
\begin{equation} \label{eq:polaron}
n_x^{\mathrm{GS}} = \ex[\big]{ b^{\dagger}(x)b(x) }_{\mathrm{GS}} = \abs[\big]{ f(x) }^2 \;,  
\end{equation}
where $f(x)$ is the Fourier transform of the variational parameters $f(p)$.  Using $f(p)=-g(p)/(c|p| + \omega_r)$, and the relationship between coupling functions and spectral density again, one finds
\begin{equation}
f(x) =\frac{-1}{\pi \sqrt{c}}\int_{0}^{\infty} d\omega\; \frac{\sqrt{J(\omega)}}{\omega+\omega_r}\cos\Bigl(\frac{\omega x}{c}\Bigr) \;. \label{fx}
\end{equation}
For the Ohmic spectral density with exponential cut-off, this expression can be evaluated analytically, yielding a slightly cumbersome expression 
that we will not reproduce here.  Like $g(x)$, this function decays as $\abs{x}^{-3/2}$ for large $\abs x$, and in \figref{fig:ohmic}(c) we see a good fit with the long-time behavior of the io-HEOM method. This confirms the spatial profile of the photon density is, in this case, dominated by the underlying spatial coupling function $g(x)$.

\subsection{Qubit energy quench and resulting emission of virtual photons}

In the field of ultra-strongly coupled cavity QED \cite{Kockum18b, forn2018ultrastrong, niemczyk2010circuit}, the question how to observe ``virtual'' ground state photons has been discussed in depth \cite{Stassi13, DeLiberato09c, de2017virtual, Anappara09, RevModPhys.84.1,DeLiberato2017} . Three common proposals include non-adiabatic quenches of the light-matter coupling strength \cite{DeLiberato09c, Gunter2009}, the use of additional uncoupled states \cite{Stassi13,Cirio16}, or the detection of virtual photons with an ancilla such as a mirror \cite{Cirio18} or an additional qubit \cite{lolli}.

Here we demonstrate an effect akin to the first of these options by suddenly raising the qubit energy to above the spectral-density cut-off, $H_S = q(t)(\omega_s /2)\sigma_z$ with
\begin{equation}
q(t) = \theta(-t+t_q) + 10 \frac{\omega_c}{\omega_s} \theta(t-t_q)
\end{equation}
where $\theta$ is the Heaviside step function, and the quench time $t_q$ is chosen to be a time where the initial transient dynamics have already subsided. A similar approach was taken in  \cite{zueco_quench}, where however the qubit was detuned above the band-gap set by the geometry of the cavity array used there.  Note that recent experiments have demonstrated directly tunable couplers, from weak to ultra-strong, in waveguide QED devices, which would also serve the same function as the quench \cite{PhysRevResearch.5.033155}.

We again employ the io-HEOM method, which supports time-dependent system Hamiltonians and time-dependent coupling strengths without any modifications, to implement this protocol. 
The results are shown in \figref{fig:ohmic_quench}.
We start with the qubit excited, and evolve until near steady-state conditions have been achieved (black curves). We then apply the quench to the qubit frequency, and observe the change in the waveguide occupation (red curves). As expected, we see a release of the virtual bound photons as new wavefronts in both directions.   A residual amount of bound-state photons remains because of the post-quench finite coupling. Experimentally, observation of the release of ground-state photons in cavities, arising from ultra-strong coupling, has proven difficult. The conditional pulse we observe in this waveguide case could be an alternative approach to achieving this goal.

\begin{figure}[!t]
\includegraphics[width=\columnwidth]{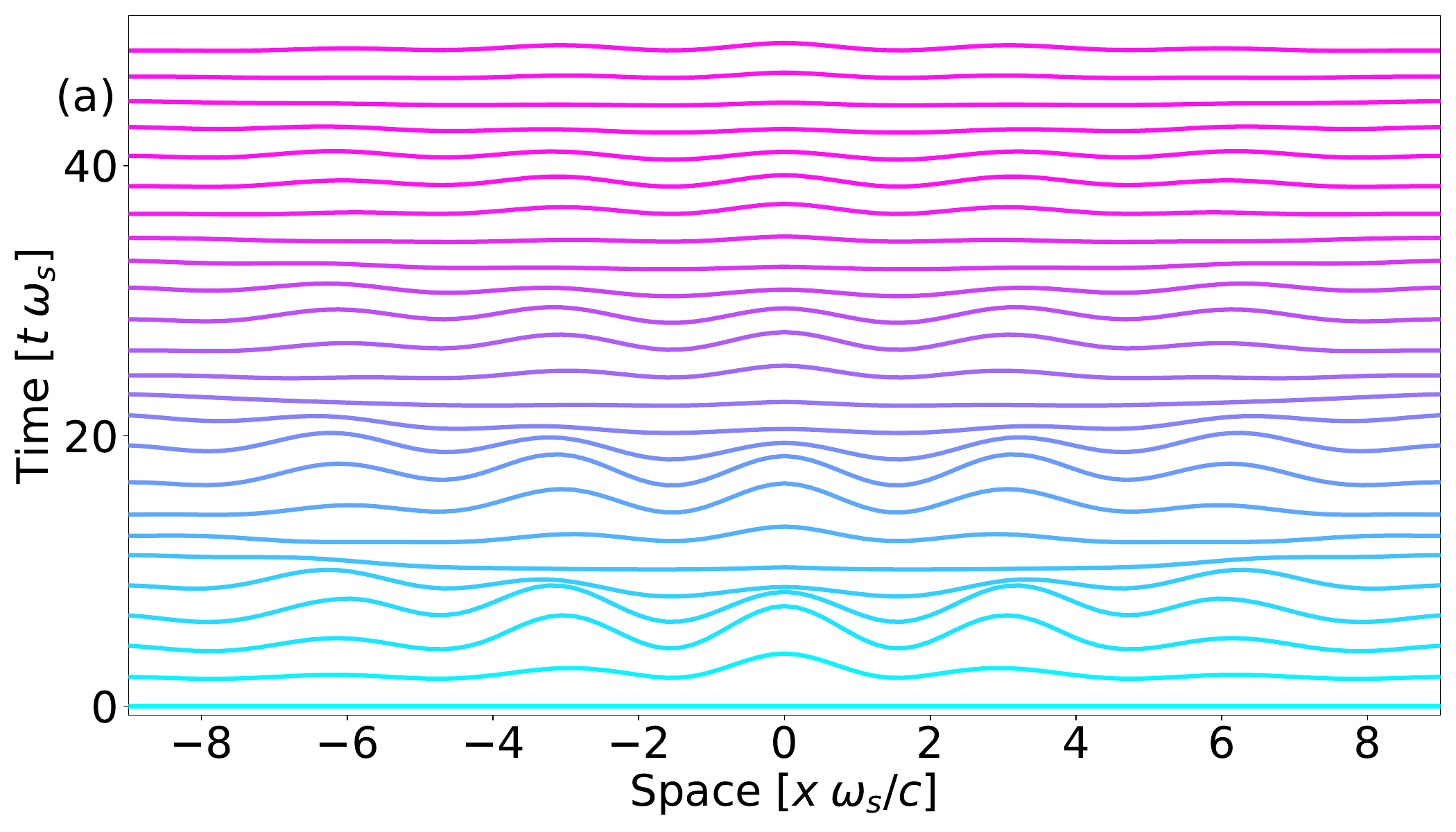}
\includegraphics[width=0.49\columnwidth]{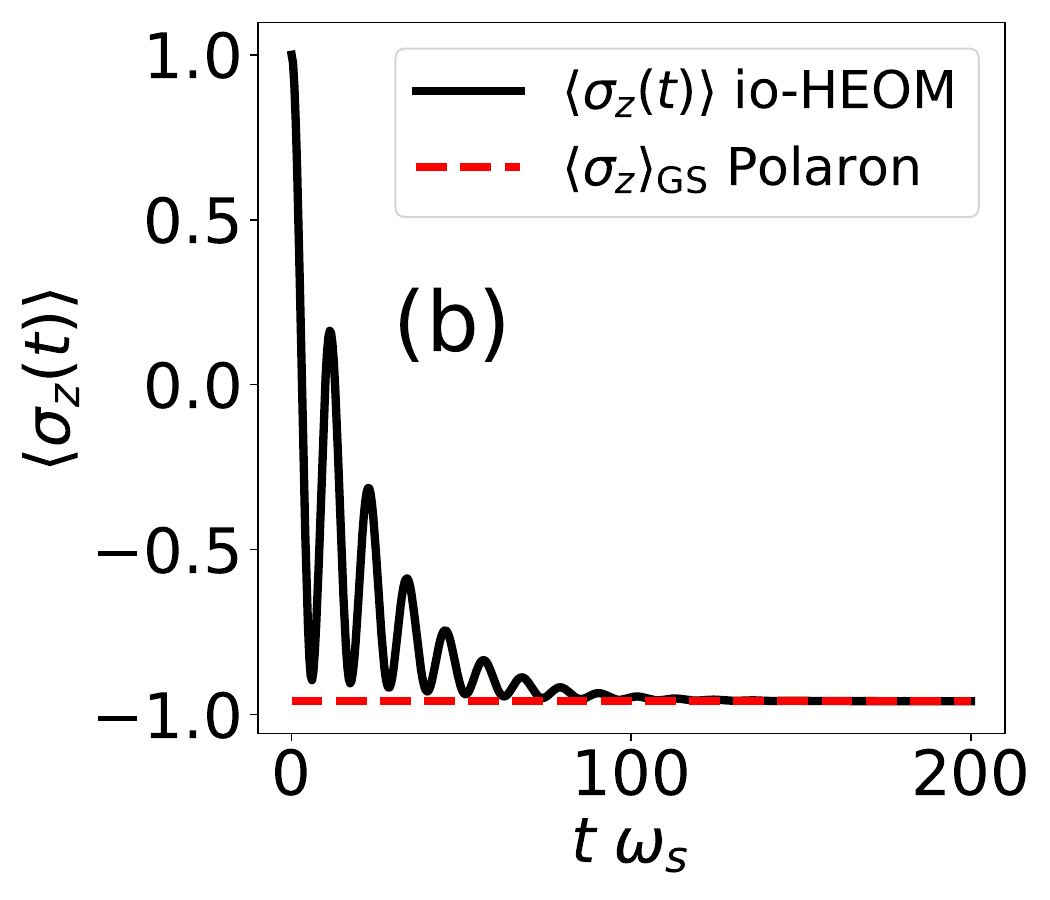}
\includegraphics[width=0.49\columnwidth]{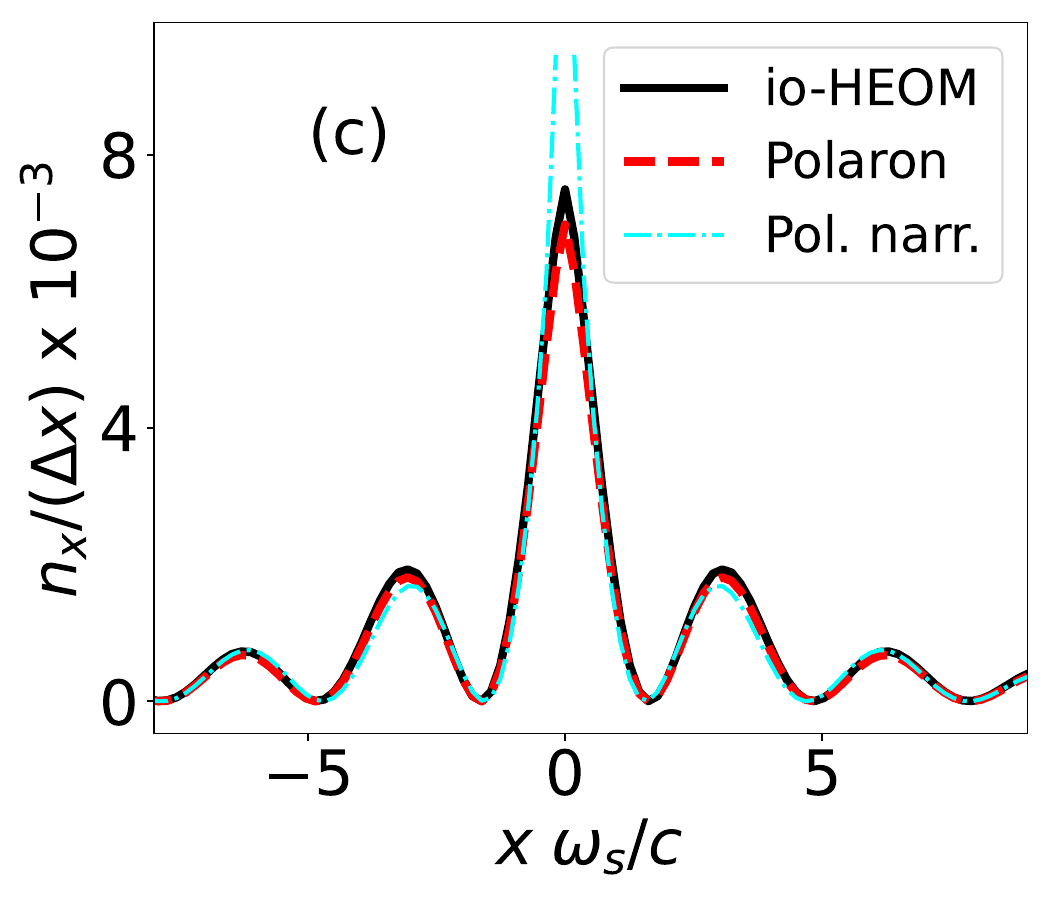}
\caption{
Dynamics and steady state for Example 2, a qubit coupled to waveguide with Lorentzian-like spectral density (due to very non-local coupling), and linear dispersion.
(a) Photon density $n_x(t)/\Delta x$ as a function of position $x \omega_s /c$, offset on the y-axis to represent different times $t \omega_s$. We use the underdamped spectral density \eqref{eq:ud_sd} with coupling strength $\lambda = 0.4 \omega_s^{3/2}$, width $\Gamma = 0.1\omega_s$, resonance frequency $\omega_0 = \omega_s$, and zero temperature ($T=0$). The initial condition is the waveguide in the vacuum and the qubit, at position $x=0$, excited. In contrast to \figref{fig:ohmic}, the wavefront now appears as a standing-wave, with peaks corresponding to the highly non-local coupling function $g(x)$ in \eqref{gxlorentz}. (b) The qubit dynamics, which are now highly non-Markovian.
(c) The long-time ($t\omega_s=200$) population of the waveguide (black curve) approaches the polaron result based on the exact spectral density (red dashed curve), and it also closely matches the polaron result based on the narrow bath approximation (cyan dot-dashed curve) except around $x=0$.  
}\label{lorentz}    
\end{figure}

\section{Example 2: Lorentzian spectral density and linear dispersion}

In our second example, our goal now is to engineer a more non-Markovian bath while remaining in the linear dispersion regime.
A commonly studied non-Markovian limit \cite{exceptional} is when the waveguide starts to act like a cavity, and exhibits intense vacuum Rabi oscillations.
The spectral density of the waveguide then has the underdamped Brownian motion form,
\begin{equation} \label{eq:ud_sd}
    J(\omega) = \frac{\lambda^{2}\Gamma\,\omega}{(\omega_0^{2}-\omega^{2})^{2}+\Gamma^{2}\omega^{2}} \;, 
\end{equation}
where $\lambda$ is the coupling strength, $\omega_0$ the resonance frequency, and $\Gamma$ the spectral width.

By imposing the spectral density \eqref{eq:ud_sd} as a condition in \eqref{jfund}, we can again find the spatial coupling function that gives rise to it.
With this spectral density, many of the appearing integrals are not analytically tractable, so we first rely on numerical integration to evaluate the correlation functions $\Omega_\pm$ that are needed for the io-HEOM, see \eqref{om_p} and \eqref{om_m}, and for the polaron calculations.

We present the results of this numerical approach in \figref{lorentz}, showing again the time evolution of the waveguide photon density and of the qubit state, as well as the long-time behavior of the photon density.   Compared to the first example, the dynamics of the qubit are much more oscillatory and non-Markovian, but the steady-state bound photon density is very small for the parameters we used.  For both qubit and waveguide, the long-time behavior qualitatively matches the polaron results.  

To gain a better analytical understanding of the coupling profile and the resulting photon density profile, we make a narrow-bath approximation ($\Gamma \ll \omega_0$) where the spectral density becomes a Lorentzian,
\begin{equation}
J(\omega) \approx \frac{\lambda^2\Gamma}{4\omega_0}\,
\frac{1}{(\omega-\omega_0)^2+(\Gamma/2)^2} \;.
\end{equation}

Using \eqref{jfund} gives the coupling function in momentum space as
\begin{equation}
g(p)
\approx
\lambda\sqrt{\frac{c\Gamma}{8\pi\omega_0}}\,
\frac{1}{\sqrt{(c\abs{p}-\omega_0)^2+(\Gamma/2)^2}} \;,
\end{equation}
and in real space as the oscillating
\begin{equation}
g(x)
\approx
\frac{\lambda}{\pi}\sqrt{\frac{\Gamma}{c\omega_0}}\,
\cos(k_0 x)\,
K_0\biggl(\frac{\Gamma \abs{x}}{2c}\biggr) \;, \label{gxlorentz}
\end{equation}
where $k_0=\omega_0/c$, and $K_0$ is the modified Bessel function of the second kind of order $0$.
The latter gives a spatial profile for large $\abs x$ of $g(x)\sim
\cos(k_0 x)\,
e^{-\Gamma \abs x/2c}/\sqrt{\abs x}$.

While the narrow-bath approximation allows us also to obtain analytical expressions for the correlation functions $\Omega_{\pm}$ in terms of Bessel functions, this approximation induces too large an error with our parameters.  However, the polaron approach continues to work well under this approximation; it evaluates to 
\begin{equation}
n_x^{\mathrm{GS}}\approx
\frac{\lambda^2\Gamma}{\pi^2 c\, \omega_0(\omega_0+\omega_r)^2}
\cos^2(k_0 x)\,
K_0^2\biggl(\frac{\Gamma \abs x}{2c}\biggr) \;, \label{pol_lorentz}
\end{equation}
where the renormalized frequency $\omega_r$ again must be determined self-consistently from \eqref{omega_r}.
Note that this result diverges at $x=0$ and, in principle, we only expect it to be a good approximation for large $\abs x$. For our parameters, as shown in \figref{lorentz}(c), we find however that it fits the exact result quite well outside the immediate vicinity of $x=0$.  In practice, the coupling function \eqref{gxlorentz} may be challenging to engineer, but demonstrates the sometimes unintuitive relationship between simple spectral densities and complex coupling functions.

\begin{figure}[t]
\includegraphics[width=\columnwidth]{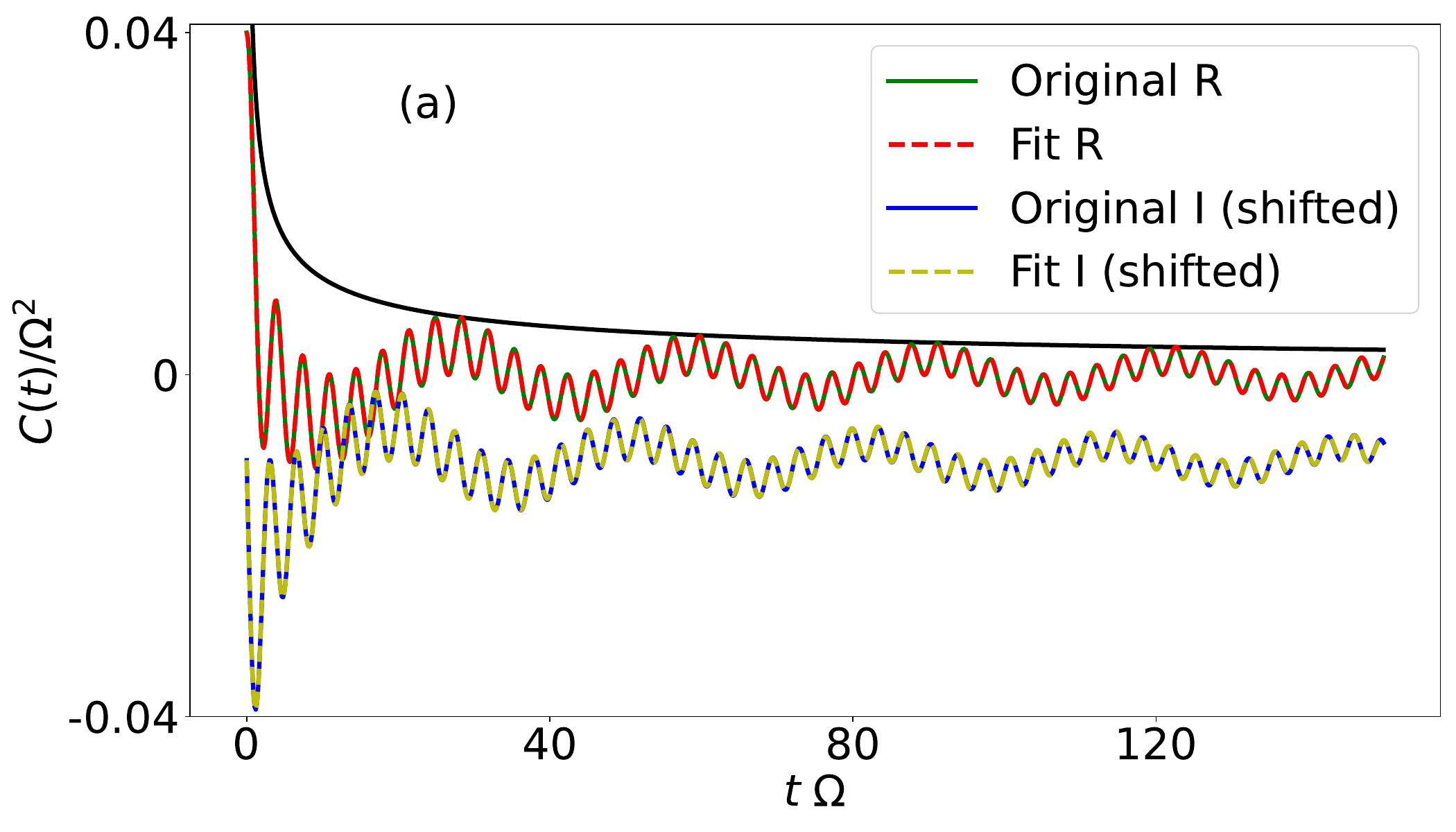}
\includegraphics[width=\columnwidth]{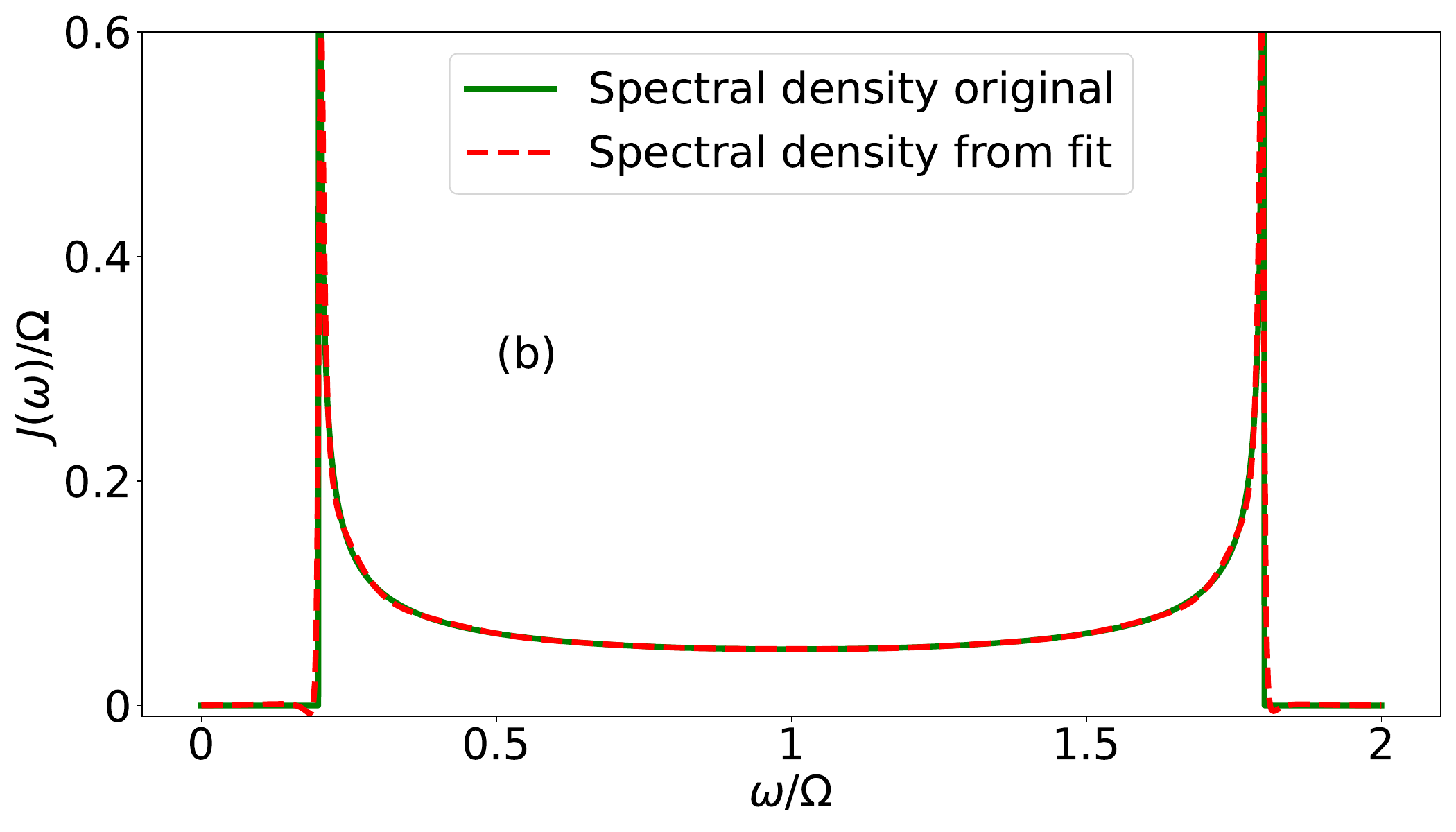}
\caption{
Correlation Function Fitting for Example 3, a cavity-array waveguide.
(a) We use a Prony method to fit the bath correlation function \eqref{cav_array_ct}. The plot shows the real (``R'') and imaginary (``I'') parts of the original correlation function (solid curves) and of the the Prony fit result using seven exponentials (dashed curves). Note that the imaginary parts have been off-set downwards by $0.01/\Omega^2$ for visual clarity.  The black solid line represents the long-time envelope of \eqref{cav_array_ct}, which is $\propto g_0^2/\sqrt{\pi g_c  t}$.
(b) Original spectral density \eqref{J_cav} (solid curve), and spectral density reconstructed from the multi-exponential approximation of the correlation function via one-side Fourier transform (dashed curve).
The system parameters used for both plots are listed in \figref{fig:cav_array_results}.
}\label{fig:cav_array_fit}    
\end{figure}
\section{Example 3: Cavity array and non-linear dispersion}

So far, we have focused on a models with linear dispersion and non-local coupling functions. However, it is also common to encounter devices where the physical structure of the waveguide implies non-linear dispersion.  As an example, we consider the widely studied cavity array model, which in one spatial dimension reads

\begin{align}
H_\text{tot} &= H_S + g_0\, \sigma_x (a_0 + a_0^{\dagger})+ \Omega \sum_{j} a_{j}^{\dagger} a_{j} \nonumber\\
&\quad + g_c \sum_{j} \bigl( a_{j}^{\dagger} a_{j+1} + a_{j+1}^{\dagger} a_{j} \bigr).
\end{align}
where $\Omega$ is the frequency of the cavities, and $g_c$ the cavity-cavity coupling.
The qubit-waveguide coupling is point-like, since the qubit only couples to the zeroth cavity with coupling strength $g_0$.
This model has the dispersion relation
\begin{equation}
\bar\omega(p) = \Omega + 2g_c  \cos(p) \;,
\end{equation}
where $p \in [-\pi, \pi)$, and the constant coupling function $g(p) = g_0 / \sqrt{2\pi}$ in momentum space.

As before, we use \eqref{jfund} to obtain the spectral density, accounting for degeneracy in the energy-dispersion relation. In the limit of an infinitely long chain, the result is
\begin{equation}
J(\omega) = J_{L}(\omega)+J_R(\omega)=\frac{g_{0}^{2}}{2g_c \,\sqrt{1-\bigl(\frac{\omega-\Omega}{2g_c }\bigr)^{2}}}\;, \label{J_cav}
\end{equation}
which has Van Hove singularities at the boundaries, i.e., at the frequencies $\omega = \Omega \pm 2g_c$ which correspond to the momenta $k=0$ and $k=\pm \pi$, see \figref{fig:cav_array_fit}(b).  The corresponding auto-correlation function at zero temperature is
\begin{equation}
C(t) = \frac{1}{\pi} \int_{\Omega-2g_c }^{\Omega+2g_c }d\omega\; J(\omega) e^{-i\omega t}= g_0^2 e^{-i \Omega t} J_0(2g_c  t) \;, \label{cav_array_ct}
\end{equation}
where $J_0(x)$ is the Bessel function of the first kind of order zero.

\begin{figure}[!ht]
\includegraphics[width=\columnwidth]{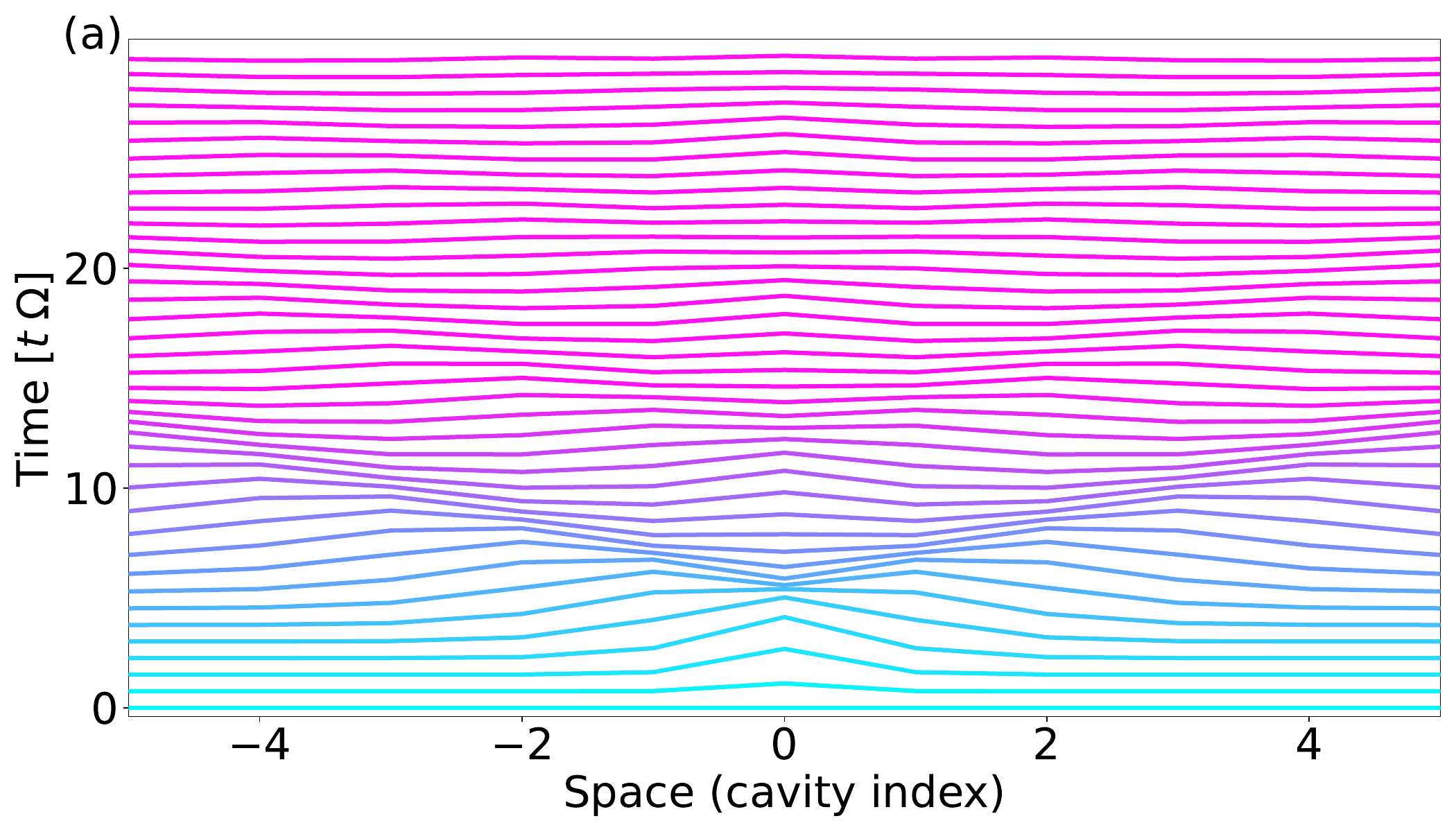}
\includegraphics[width=\columnwidth]{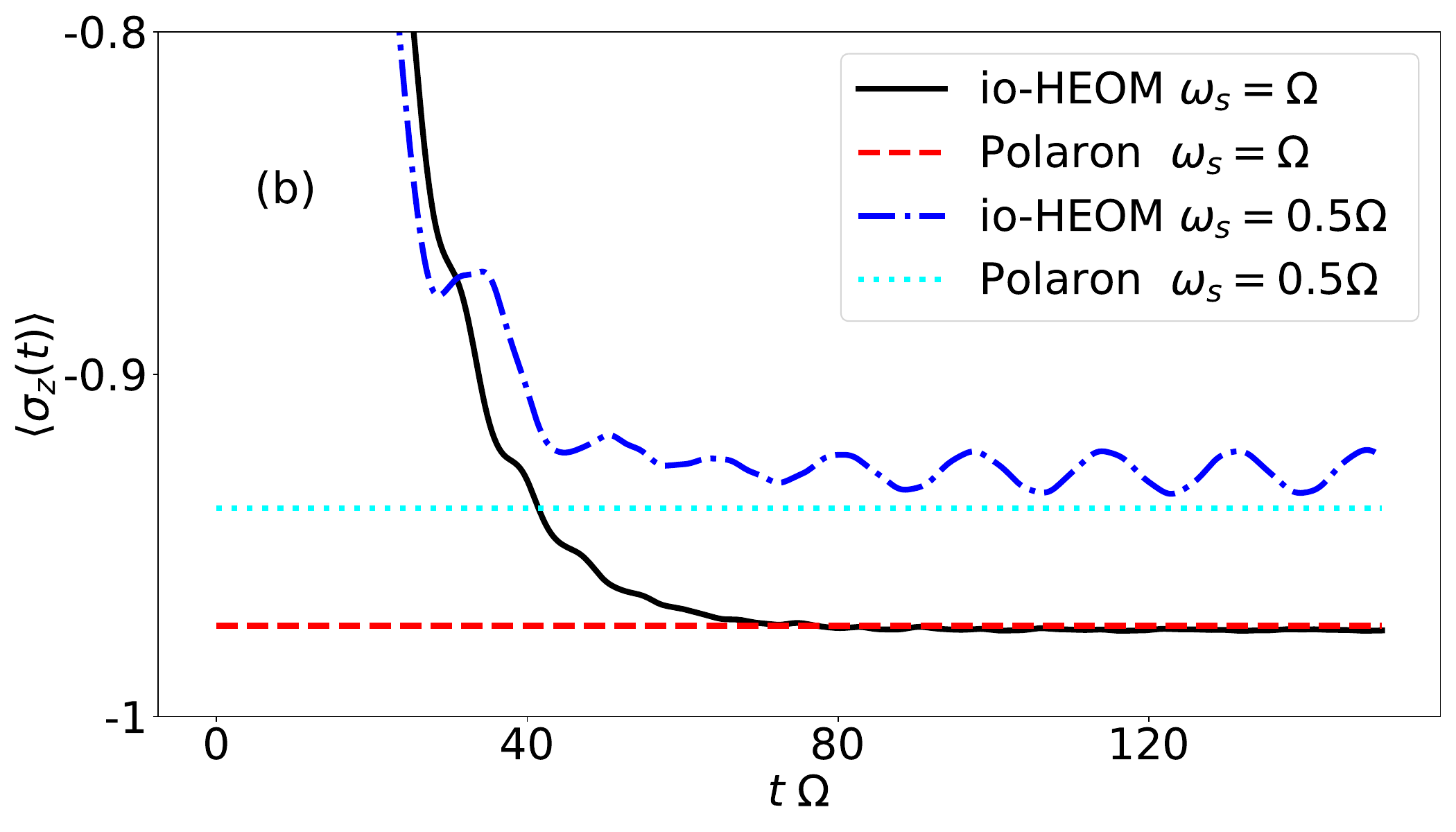}
\includegraphics[width=\columnwidth]{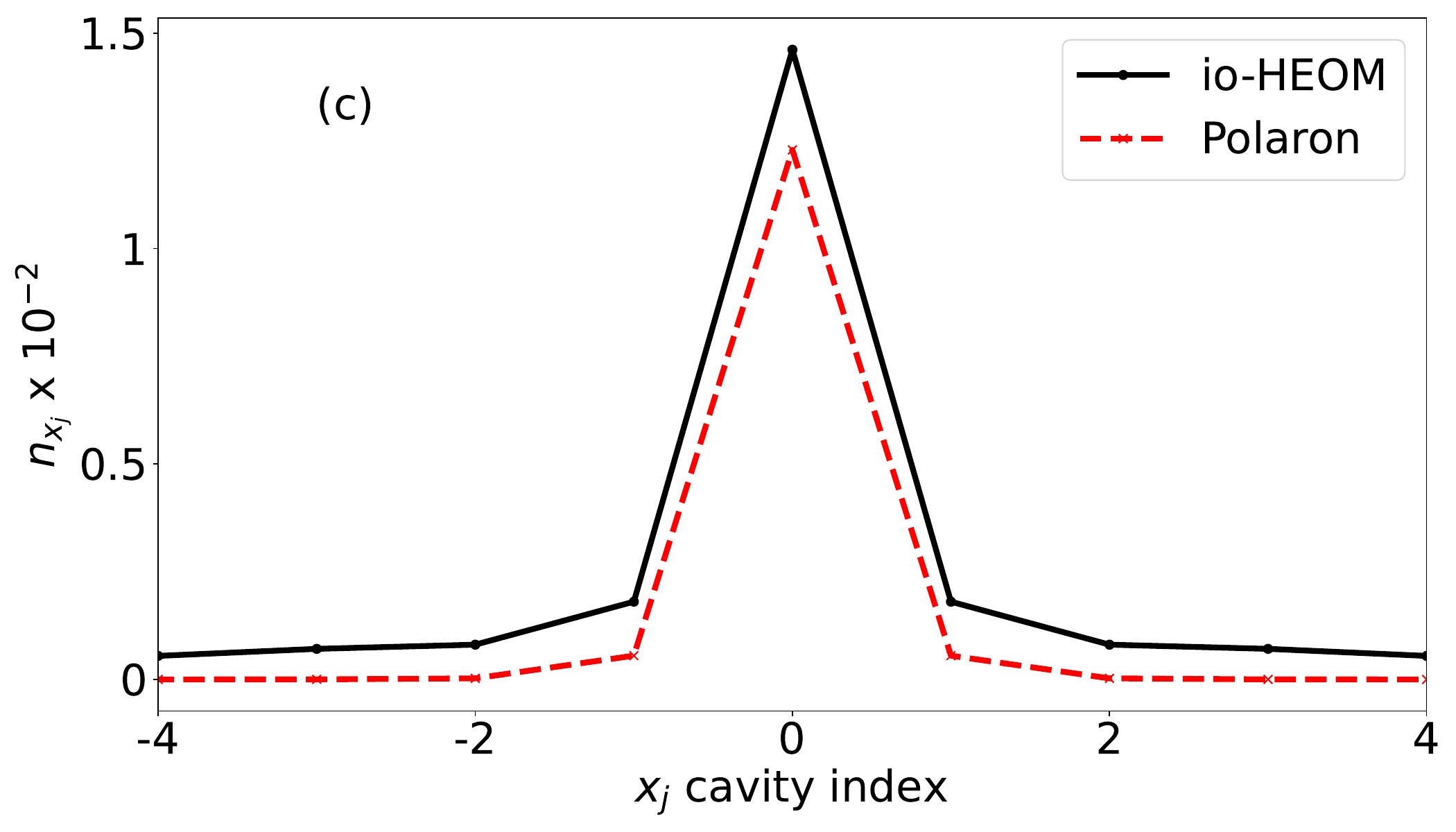}
\caption{
Dynamics and steady state for Example 3, a qubit coupled to a cavity-array waveguide with very non-linear dispersion.
(a) Photon occupation as a function of the discrete cavity index $m$, offset on the y-axis to represent different times $t \omega_s$. We used the qubit-cavity coupling $g_0 = 0.2 \Omega$, the cavity-cavity coupling $g_c  = 0.4 \Omega$, and zero temperature ($T=0$). The cavity frequency is chosen as $\omega_s=\Omega$, so that the qubit sits in the middle of the spectral density (see \figref{fig:cav_array_fit}). The initial condition is the cavity-array in the vacuum and the qubit, coupled to the $m=0$ cavity, excited.  As before, we see wave fronts propagating in both directions and a residual bound photon occupation around the qubit. Due to the slow decay of the correlation functions, the cavity populations here exhibit persistent oscillations. We can approximately reproduce both (b) the residual qubit population and (c) the waveguide spatial profile at long but finite times $t\Omega = 150$ with Silbey-Harris polaron theory. In (b), we additionally show the shifted qubit frequency $\omega_s=0.5 \Omega$, which is closer to the singularity in the spectral density. This induces larger coupling and hybridization and even more persistent oscillations.}\label{fig:cav_array_results}    
\end{figure}

This result shows that the correlation function contains a slow-decaying $1/\sqrt{t}$ component. The standard formulation of the HEOM, which we introduced above, requires a multi-exponential approximation of the correlation function, as in \eqref{ansatz}. It is challenging to capture the slowly decaying correlation functions across all time scales with a multi-exponential ansatz. Intriguingly, a variant of HEOM using a Bessel function ansatz has been proposed in \cite{heomengel}. 
However, in this paper, we choose as our strategy to retain the exponential ansatz and fit the correlation functions up to a fixed finite time, and evolve the equations of motion only up to that time. Hence, correctness beyond this time scale is not guaranteed \cite{plenio}.  In practice, we found that the Prony method produces good fits for this correlation function, see \figref{fig:cav_array_fit}.

Despite the more complex dispersion relation for this example, the correlation functions between output and coupling terms can still be evaluated analytically:
\begin{equation}
\Omega_+  = (-i)^m g_0\, e^{i\Omega(t_\text{out}-t)} J_m\bigl[2g_c (t-t_{out})\bigr]
\end{equation}
and $\Omega_- = \Omega_+^\ast$.
Here, $m = \operatorname{int}( x_\text{out} / \Delta x )$ is the integer site index, $\Delta x$ the array spacing, and $J_m$ the Bessel function of the first kind of order $m$.  Using this result with the io-HEOM method, we simulate the cavity-array population, shown in \figref{fig:cav_array_results}. As with the continuous waveguide examples, we plot the populations as slices at different times, starting with an initially excited qubit and the cavity array in its vacuum state. 

Like for the continuous waveguide, there is an initial emission into the array, and residual bound photons around the qubit in the steady state. However, there is a strong difference in the dynamics: the populations of the cavities continue to oscillate for long time scales, as do the system observables, and there is still a distinct difference between the long-time HEOM result up to $t\Omega = 150$ and the ground-state polaron prediction, which we attribute to not having reached the steady-state yet. Note that this is still an infinite chain, of which we plot only a small section, but the slow decay is reflecting the strong non-Markovianity of this challenging case.  

The polaron result (see \cite{zueco_quench} for a detailed derivation for the cavity array case) exhibits exponential dependence on the cavity site $m$, following $n_m^{\mathrm{GS}} \propto \exp\{-\abs{m}/\xi\}$ with $\xi^{-1} \propto \mathrm{arccosh}[(\Omega+\omega_r)/2g_c]$ instead of the power law that we saw in previous examples. Since the coupling is now point-like, this dependence arises from the cavity-array properties itself, with a localization length related to distance to the singularities in the spectral density.

In \figref{fig:cav_array_results}, we first set the qubit energy in the middle of the spectral density shown in \figref{fig:cav_array_fit}(b). If we move the qubit energy towards the singularities in the spectral density, see \figref{fig:cav_array_results}(b), we see a much larger influence of the array correlation functions on the qubit dynamics; the qubit population exhibits large persistent oscillations. Propagating the HEOM very close to the singularity becomes challenging, as it requires both accurate fits of the free-bath correlation functions and a large cutoff of the hierarchy equations themselves, due to the effectively stronger coupling. Propagating the HEOM for this parameter choice until the steady-state is thus not feasible. As mentioned earlier, adopting alternative HEOM formulations relying on a Bessel function ansatz may be a powerful way to fully capture this challenging example on all time scales \cite{heomengel}.

\section{Conclusions}

The use of structured non-perturbative and non-Markovian waveguides  in quantum science is growing rapidly. They are an important tool in both ``analog'' fundamental physics, like test-beds for many-body spin-boson physics \cite{magazzu2018probing, Puertas, entharve, gong}, and for quantum information science (for enhanced qubit readout, long-distance interactions \cite{PhysRevLett.120.140404,      zueco_quench}, state transfer, and more).  

Here we extended the recently introduced input-output HEOM framework \cite{ioheom} to deal with these challenging systems. We demonstrated that the framework can easily, and robustly, deal with non-Markovianity arising from both spatially extended coupling functions and from non-linear dispersion. In the first case, for non-local coupling, we demonstrated how an Ohmic bath with exponential cut-off can be ``engineered'' from a power-law continuous coupling function, and gives rise to a large non-perturbative bound photon density in the waveguide. We also demonstrated a simple protocol \cite{zueco_quench} for releasing these photons, and how a more complex coupling function can give rise to a more dynamically non-Markovian Lorentzian-like bath.  

In the second case, for non-linear dispersion, we showed how the highly oscillatory and slow power-law decay of the cavity-array correlation functions gives rise to persistent system and waveguide oscillations.  Code examples, using an augmented version of the HEOM solver in QuTiP \cite{Qutip1, Qutip2, Qutip3}, are made available as supplementary information.

Exciting future work that remains to be done includes going beyond one-dimensional systems, introducing topological effects, studying entanglement harvesting \cite{entharve, entharv2, entharv3} and using input-pulse/scattering to probe bound-states \cite{zuecodynamicbound}. 

\acknowledgments
We thank Mauro Cirio and Peng-fei Liang for discussions and feedback, and Mana Lambert for designing \figref{schematic}.  
N.L. is supported by MEXT KAKENHI (Grant No. JP24H00816 and Grant No. JP24H00820).
F.N. is supported in part by the Japan Science and Technology Agency (JST)
[via the CREST Quantum Frontiers program Grant No. JPMJCR24I2,
the Quantum Leap Flagship Program (Q-LEAP), the Moonshot R\&D Grant Number JPMJMS256E,
and the ASPIRE program (Grant Number JPMJAP2513)]
and the Office of Naval Research (ONR) Global (via Grant No. N62909-23-1-2074).
YNC acknowledges the support of the National Center for Theoretical Sciences and the National Science and Technology Council, Taiwan (NSTC Grant No. NSTC 114-2112-M-006-015-MY3).

\appendix
\makeatletter
\renewcommand{\p@subsection}{\Alph{section}}
\makeatother

\section{HEOM} \label{app:heom}

In the context of generic open quantum systems, the HEOM method is a commonly used numerically exact tool to solve a discrete quantum system in contact with a (Gaussian) bosonic environment, whose total Hamiltonian takes the form
\begin{equation}
H_{\mathrm{tot}} = H_S + H_{\mathrm{bath}} + H_{\mathrm{int}} \;.
\end{equation}
The HEOM relies on a linear coupling $H_{\mathrm{int}} = S X$, where $S $ is some arbitrary system operator, and $X$ a bath operator linear in the bath modes, $X = \sum_k f_k (a_k+a_k^{\dagger})$. Its evolution under $H_{\mathrm{bath}}$ alone preserves linearity and Gaussianity.  Given the standard definition of the spectral density 
\begin{equation} \label{ct}
J(\omega) = \pi\sum_k \abs{f_k}^2 \delta(\omega -\omega_k) \;,
\end{equation}
we define the free-bath correlation function 
\begin{align}
C(t) &= \ex[\big]{ X(t)X(0) } \nonumber\\
&= \int_0^{\infty} d\omega\; \frac{J(\omega)}{\pi} \bigl[ \coth(\beta\omega/2) \cos(\omega t) - i \sin(\omega t) \bigr] \;, 
\end{align}
where $\beta$ is the inverse bath temperature.

The HEOM assumes an ansatz of 
\begin{align}
C(t) &= C_R(t) + i C_I(t), \nonumber\\
C_{R/I}(t) &= \sum_{k=1}^{N_{R/I}} c_k^{R/I} \exp\bigl(- \gamma_k^{R/I} t \bigr) \;, \label{ansatz}
\end{align}
where $c_k^{R/I}$ and $\gamma_k^{R/I}$ are arbitrary complex parameters. While the HEOM is in principle exact, the number of exponents $N_{R/I}$ is a convergence parameter in terms of how well this ansatz converges to that of the true bath correlation function.  Interestingly, this ansatz is common for other numerical techniques, and a great deal of effort has been made in optimizing the number of exponentials needed, since the numerical cost of the HEOM scales with the number of exponents.

The formal equations of motion of standard HEOM, given the above definitions, can be written (in one of many equivalent ways) as, 
\begin{align}
\dot{\rho}^{(\bar{n})}(t)
&= -i[H_S, \rho^{(\bar{n})}(t)] -  \sum_{k}\sum_{j=R/I}n_{jk} \gamma_k^{j} \rho^{(\bar{n})}(t) \nonumber\\
&\quad - i\sum_k c_k^R \mathcal{S}^{\times} \rho^{(\bar{n}_{Rk}-1)}(t) + \sum_k c_k^I \mathcal{S}^{\circ} \rho^{(\bar{n}_{Ik}-1)}(t)\nonumber\\
&\quad - i\sum_{k}\sum_{j=R/I}\mathcal{S}^{\times}\rho^{(\bar{n}_{jk}+1)}(t)\nonumber\\
&\equiv \mathcal{M}_{\mathrm{HEOM}}\Bigl[\rho^{(\bar{n})}(t)\Bigr] \;, \label{heomeom}
\end{align}
where $\mathcal{S}^{\times} = S[\cdot] - [\cdot]S$ and $\mathcal{S}^{\circ} = S[\cdot] + [\cdot]S$, and $\bar{n} = (n_{R1},n_{R2}, \cdots, n_{I1}, n_{I2}, \cdots)$ is a multi-index of integers tracking the system and auxiliary density operators. Each index takes values $n_{jk} \in {0 \dots N_c}$, where the hierarchy cutoff $N_c$ is an additional truncation parameter (in addition to the fitting truncation parameters $N_{R/I}$).  The notation $\rho^{(\bar{n}_{jk}\pm 1)}$  means we increase or decrease the entry $n_{jk}$ of the multi-index by one.

The state $\rho^{(0)}(t)$ is the exact system state, while the  other auxiliary density operators ($\bar{n}\neq 0$) encode the bath dynamics.  These can be understood as being equivalent to system states conditioned on projections of purified pseudomodes onto a particular state \cite{liang2024purifiedinputoutputpseudomodemodel}.

\section{Polaron method for the continous waveguide}

Here we give the detailed steps for the results given in \eqref{omega_r} and \eqref{fx}.  Following \cite{zueco_quench, Silbey}, we first apply the unitary transformation
\begin{equation}
U_p = \exp(\sigma_x A)
\end{equation}
with
\begin{equation}
A = \int dp\; f(p) \bigl[ b^{\dagger}(p) - b(p) \bigr] \;,
\end{equation}
where $f(p)$ is a function which needs to be determined variationally. We assume it is real valued here, as is the case for all the examples we explore, and to produce more compact formulas below. Noting that $[b(p), A] = f(p)$, 
\begin{equation} 
U^{\dagger} b(p) U = b(p) + \sigma_x f(p)
\end{equation} 
and 
\begin{equation}
    U^{\dagger} b^{\dagger}(p) U = b^{\dagger}(p) + \sigma_x f(p),
    \end{equation}
    and applying this unitary to the Hamiltonian \eqref{eq:1D} leads to,
\begin{align}
\bar{H}_\text{tot} &= U^{\dagger}H_\text{tot}U \nonumber\\
&=
\underbrace{\int dp\; \bar\omega(p)\,b^\dagger(p)b(p)}_{\text{free bath}}
+
\underbrace{\vphantom{\int}\frac{\omega_s}{2}\Bigl(\sigma_x^+e^{-2A}+\sigma_x^-e^{2A}\Bigr)}_{\mathclap{\text{tunneling dressed by displacement}}}\nonumber\\
&\quad +
\underbrace{\sigma_x\int dp\; \bigl[\bigl\{\bar\omega(p)f(p)+g(p)\bigr\}\bigl\{b(p)+b^\dagger(p)\bigr\}\bigr]}_{\text{residual linear coupling}}\nonumber\\
&\quad +
\underbrace{\int dp\;\bigl[\bar\omega(p)f(p)^2+2f(p)g(p)\bigr]}_{\text{energy shift (c-number)}} \label{Hbar}
\end{align}
Here, we have defined $\sigma_x^{+}= \ket{+}\bra{-}$ and $\sigma_x^{-}= \ket{-}\bra{+}$, i.e., raising and lowering operators in the eigenbasis of $\sigma_x$. One then introduces a ground-state ansatz $\ket{\Psi_{GS}} = U \ket{0,s}$, where $\ket{0}$ is the vacuum state for the waveguide and $\ket{s}$ is an arbitrary qubit state.  This ansatz is exact when $\omega_s \rightarrow 0$, whence $f(p) = -g(p)/\bar\omega(p)$. 

Away from this limit, we can find an approximate ground-state by minimizing the energy  $E_{GS}= \bra{s,0}\bar{H}_\text{tot}\ket{0,s}$.  One proceeds by neglecting the linear coupling term, and approximating the dressed tunneling term with the mean field 
\begin{equation}
B_{\pm} = \bra{0}\exp(\pm 2A)\ket{0} = \exp \biggl[ -2 \int dp\; f(p)^2\biggr] \;,
\end{equation}
which leads to an effective spin Hamiltonian
\begin{equation}
\bar{H}_S = \frac{\omega_s}{2} \exp\left[-2 \int dp\; f(p)^2\right]\sigma_z = \frac{\omega_r}{2} \sigma_z \;. \label{Hbars}
\end{equation}
In minimizing $E_G$  with $\delta E_G/\delta f(p) =0 $, we simultaneously minimize this dressing and the final energy shift term in \eqref{Hbar}, 
\begin{align}
(\bar\omega(p) +\omega_r)f(p) + g(p) = 0 \\
f(p) = \frac{- g(p)}{\bar\omega(p) + \omega_r} \;.
\end{align}
Now we can finally define the renormalized qubit frequency by substituting this result back into $\omega_r$, and using $J(\omega) = \pi \int dp\; g(p)^2 \delta[\omega-\bar\omega(p)]$,
\begin{equation}
\omega_r = \omega_s \exp\biggl[-\frac{2}{\pi}\int_0^{\infty} d\omega\;  \frac{J(\omega)}{(\omega+\omega_r)^2}\biggr] 
\end{equation}
We can then, given $J(\omega)$, solve for $\omega_r$ self-consistently. 

As in the example in \cite{zueco_quench}, the qubit expectation value of the ground-state polaron approximation can be found to be given simply by 
\begin{align}
\bra{-1,0}U^{\dagger} \sigma_z U\ket{-1,0} &= \bra{-1,0}B_-\sigma_x^{+} + B_+\sigma_x^{-}\ket{-1,0} \nonumber\\
&= \frac{\omega_r}{\omega_s}\bra{-1}\sigma_z\ket{-1} = -\frac{\omega_r}{\omega_s} \;.  
\end{align}
How to extract the polaron ground state density $n_x^{\mathrm{GS}}$ was demonstrated in the main text, see \eqref{eq:polaron} and \eqref{fx}.

\bibliography{refs}

\end{document}